# Spectral pulsations of dissipative solitons in ultrafast fiber lasers: period doubling and beyond

*Zhiqiang Wang[1,3], Aurélien Coillet[2], Saïd Hamdi[2], Zuxing Zhang[3,*] and Philippe Grelu[2,*]*

[1] Aston Institute of Photonic Technologies, Aston University, Birmingham B4 7ET, UK
[2] Laboratoire ICB UMR 6303 CNRS, Université Bourgogne Franche-Comté, 9 avenue Alain Savary, F-21000 Dijon, France
[3] Advanced Photonic Technology Lab, College of Electronic and Optical Engineering, Nanjing University of Posts and Telecommunications, Nanjing 210023, China

Corresponding author(s): zxzhang@njupt.edu.cn; philippe.grelu@u-bourgogne.fr

Abstract: Period doubling is a universal bifurcation of central importance in all disciplines of nonlinear science, which generally signals the existence of chaotic dynamics in the vicinity of the system parameters. Although observed in diverse ultrafast laser configurations, there is still no consensus on its physical origin. The observations also include other types of pulsating dissipative solitons, with either short or long periods. Real time spectral characterization allows to investigate optical spectral oscillations, whose features reveal the intracavity dynamics leading to instabilities. Following a contextual review, this article presents a variety of period doubling dynamics manifesting in the spectral domain of dissipative solitons. These dynamics are obtained with ultrafast fiber lasers featuring either anomalous or normal dispersion. It reveals a sequence of period doubling bifurcations and instabilities within transient dynamics, unveiling intertwined bifurcations and the entrainment of new pulsating frequencies. The oscillating frequencies tend to lock to the integral of roundtrip numbers as well as coexist with period doubling, demonstrating new combinations of the period doubling bifurcation with other bifurcations. These experimental findings are confirmed by numerical simulations, emphasizing both the universality of the period doubling bifurcations and their potentially highly complicated manifestations within ultrafast laser systems.

## 1. Introduction

By construct, in the scheme of passive mode locking, an intracavity laser pulse experiences a periodic propagation medium. However, even if the parameters of the propagation medium are deemed perfectly periodic, there is still a chance that the output pulse train will not adopt the fundamental cavity periodicity, even in the absence of noise considerations. This is deeply rooted into nonlinear dynamics, which allows several types of bifurcations. A bifurcation occurs when a small change in a physical parameter produces a major change in the organization of the system. In periodic systems, the period-doubling bifurcation is of central importance. Period doubling takes place when a slight change in the system parameters leads to the emergence of a new periodic orbit that doubles the period of the original orbit. Being prevalent in all disciplines involving nonlinear dynamics, period doubling is a universal phenomenon: it is widely reported in fluid dynamics,[1,2] ecosystems,[3,4] biology,[5–7] and nonlinear optical dissipative systems.[8–14] Furthermore, the cascade of period-doubling bifurcations represents a well-known route to chaotic dynamics. Therefore, understanding the dynamics of period-doubling and its subsequent destabilizations provides an essential tool for the monitoring of nonlinear systems. It also allows developing early warnings for applications requiring a stable periodic operation.

Ultrafast lasers are nonlinear dissipative oscillators with a virtually infinite number of degrees of freedom – namely, the field amplitude values as a function of time. They make a convenient experimental platform for studying period doubling as well as more complex bifurcations.[15–21] Following a bifurcation that generates a periodic oscillation of the pulse parameters when monitored at a fixed location – such as the laser output port – the laser dynamics is qualified as a pulsating dissipative soliton regime.[22–24] Pulsating solitons can involve short-period pulsations (SPP) whose periodicity remains comparable to the cavity roundtrip time, long-period pulsations (LPP), or even multiple-period pulsations that combine both.[16] We could initially think that LPP would build up from a cascade of period-doubling bifurcations. However, this is not the usual situation, as cascaded bifurcations tend to shortly lead to chaotic dynamics, whereas stable LPP can be routinely observed in ultrafast fiber laser dynamics; instead, they result from Hopf-type bifurcations yielding a limit-cycle attractor where the pulse features oscillate over successive roundtrips, with a periodicity that is typically in the range of 10 to 50 cavity roundtrips.[16,25,26] In this way, the periodicity of LPP no longer needs to remain commensurate with the cavity roundtrip periodicity.

In ultrafast laser experiments, LPPs often manifest when the pump power is reduced from the region of stationary mode locking,[25,27] whereas SPPs usually occur beyond the stationary mode locking region.[16,28] This marked difference indicates different bifurcation mechanisms involved in either LPP or SPP. However, the existence of multiple-period pulsations combining LPP and SPP, demonstrated both experimentally and numerically,[16] reveals the existence of complex intertwined LPP and SPP bifurcation regions. Other bifurcations, such as period-3, period-6 and period-N as well as the transition from period-1 to period-3 or to period-4 without period doubling are also observed.[16,29]

We note that in recent literature concerning ultrafast fiber laser dynamics, pulsating soliton dynamics with LPP have sometimes been called “breathing soliton dynamics”.[25,26,30] To our opinion, the later terminology should be handled with care. First, the notion of breather solitons originates from the Hamiltonian dynamics that takes place in passive and conservative systems such as the lossless, translationally-invariant, optical fiber modelled by the nonlinear Schrödinger equation (NLSE). In the latter, breather solitons arise from the interplay between Kerr nonlinearity and chromatic dispersion effects, and manifests as a pulse that oscillates on a nonzero background.[31,32] If we move out of Hamiltonian dynamics, the closest photonic system where the breather soliton concept can be extended is probably the driven-cavity resonator, where cavity solitons also sit on a background, and where a high-finesse nonlinear cavity can be modeled by the distributed Lugiato-Lefever equation.[33,34] However, when we translate to ultrafast fiber lasers, the interplay of major nonlinear dissipative effects (from saturable absorption and gain saturation, notably) will provide distinct dynamical features, such as the very existence of bifurcations, attractors, and the disappearance of the strong continuous background in most of the pulsating dynamics, in contrast to the original attributes of breather solitons.

Moreover, ultrafast laser dynamics also feature complex bifurcations leading to non-periodic and chaotic pulsations, such as in the so-called quasi-periodic “soliton explosions”.[22,30,35] Abrupt bifurcations of a different nature can also take place in ultrafast lasers, such as the transition to a multiple-pulse regime or to an incoherent localized pulse structure. The multi-pulsing instability is indeed a recurrent phenomenon in most ultrafast lasers that schematically originates from the overdrive of nonlinearities when the pumping power is increased. Furthermore, real-time experimental observations on the dynamics of multi-pulse interactions and on the build-up dynamics of incoherent dissipative solitons in ultrafast fiber lasers have shown a link between the pulsating solitons and the onset of chaotic waves.[30,36]

Nevertheless, when moving the laser parameters around from the domain where stable stationary pulses are obtained, the period-doubling bifurcation often constitutes the primary bifurcation that occurs before more complex bifurcations and chaotic pulse evolution manifest. This bifurcation is either abrupt (sub-critical) or smooth (supercritical). Let us also recall that for special parameters sets, a period-tripling bifurcation can take place instead, as a primary destabilizing bifurcation, along with numerous other possibilities that are attributes to complex nonlinear dissipative systems having many degrees of freedom.[16]

With the previous considerations in mind, it is instructive to inquire about the influence of the chromatic dispersion regime on the bifurcations that lead to SPP or LPP pulsed laser regimes. In Hamiltonian scalar photonic systems, the dispersion sign is decisive about the existence of solitary wave structures and dynamics. Bright solitons and associated breathers require an anomalous dispersion to exist. This is no longer the case in dissipative systems: the presence of nonlinear dissipation enables the formation of stable bright solitary waves in the normal dispersion regime, within broad ranges of laser parameters.[15,37–40] Dissipative solitons are indeed solitary waves that result from an average balancing between dispersive effects on the first hand, and between dissipative effects on the second hand. These two balances are intimately coupled, therefore the interplay between dissipative effects (gain and loss) significantly complexifies the investigation and mapping of the nonlinear dynamics, while yielding abundant new solitary waves solutions. By extension, whereas we anticipate that chromatic dispersion remains a major parameter in the bifurcation dynamics of ultrafast lasers, it should not rule alone the very existence of these bifurcations.

The period doubling of dissipative solitons is an example of such dynamics where the destabilization of the delicate equilibria between dispersive and dissipative effects leads to complex dynamics. This phenomenon has been widely inspected numerically using the cubic-quintic complex Ginsburg Landau equation (CQGLE), which is a central model for dissipative optical solitons.[22,23,41,42] In the experimental area of ultrafast lasers, after a wake of early research investigations,[16,43–45] the experimental study of the period doubling dynamics has been revived recently,[25,27,28,46–50] thanks to the popularization of the dispersive Fourier-transform (DFT) technique.[51] DFT is a spectral characterization method allowing to capture the spectral intensity profile of pulses in the time domain over consecutive cavity roundtrips at up to GHz repetition rates. The inception of DFT actually originates from the early 1970's.[52] By recording successive optical spectra with an appropriate spectral resolution, DFT enables the investigation of new types of pulsating solitons that are otherwise not visible from the simple monitoring of the energy of output pulses[28] and makes it possible to compare numerical simulation and experimental recordings in a more convincing way.

In the following of this article, we present a collection of experimental observations of period-2 pulsating solitons in ultrafast fiber lasers operated under distinct dispersion regimes and different operating wavelengths, thus confirming the prominence and universality of the period doubling bifurcation within a broad class of ultrafast lasers. We employ real-time DFT measurements to investigate the transition from the stable mode locking regime to period-2 and then to (LPP) period-N pulsations. The instabilities arising during the transitions are analyzed, revealing the generation of new frequencies during the sequence of bifurcations. We report on an original dynamical locking phenomenon, where the periodicity of the oscillations tends to lock on multiples of the cavity roundtrip. Numerical simulations based on a parameter-managed propagation model reproduce well the experimental observations. These simulations demonstrate that the entrainment of the pulsation frequencies during the transition take root from the oscillating behavior of the soliton optical spectrum over different periodic trajectories, which yields the synchronization of the entrained frequencies to an integer number of cavity roundtrips.

## 2. Period doubling in the anomalous average dispersion regime

### 2.1. Stable period-2 solitons

We first studied the generation of period-2 solitons in a fiber laser operating in the anomalous path-averaged dispersion that emits short pulses in the 1.5-1.6 μm wavelength region. The laser gain is provided by a short length of erbium-doped fiber (EDF). The normal dispersion of the latter is overcompensated by the anomalous dispersion of the passive fibers (SMF). Mode locking is achieved through a virtual saturable absorption based on the nonlinear polarization evolution (NPE) that takes place in the optical fibers, followed by discrimination by an intracavity polarizing beam splitter. More details of the experimental and measurement setup are provided in the experimental section. With a pump power of about 80 mW and by adjusting the polarization controller, thus fine-tuning the virtual saturable absorption, we obtain a stable single-pulse mode-locked regime with constant pulse energy over consecutive roundtrips. However, the DFT signal readout on the oscilloscope after propagation through the 1.3-km long dispersion compensation fiber (DCF) is not constant. This illustrates the existence of "invisible pulsations",[28] namely, pulsations that do not entail any obvious change in the pulse energy but nevertheless involve a significant pulse reshaping over successive roundtrips. This contrasts with the early period-2 observations in ultrafast lasers and driven cavities, which relied on the energy monitoring and could not detect such pulsations.[11,16] From the DFT measurement, we observe a period of two roundtrips for the soliton spectrum, as shown in Fig. 1 (a) & (b). One can notice that the oscillation is mainly manifested within a relatively narrow band around the center of the soliton spectrum; on the time-averaged spectrum recorded by an optical spectrum analyzer (OSA), such dynamics leaves a tiny bump as a hint. The Fourier transform of single-shot DFT spectra yields first-order temporal autocorrelation (AC) traces. In direct correspondence with the optical spectra, the AC traces indicate slight deformations from an ideal bell-shaped soliton, with extended tails bearing faint oscillations (See in Fig. 1(c)) that emanates from the dispersive waves that are radiated out periodically from the soliton bulk. Following the adjustment of the polarization controller, another example of period-2 soliton dynamics with enhanced oscillations is provided in the Supplementary Material section (See

Fig. S1). Our numerical simulations reproduce well the experimental observations. The Simulation Model section of the Supplementary Material provides details of the numerical model and parameters used. When the small signal gain parameter is set at 16 $m^{-1}$, the simulation results displayed in Fig. 1(d-f) clearly show the similar period-2 dynamics of solitons in the anomalous-dispersion regime of the ultrafast laser.

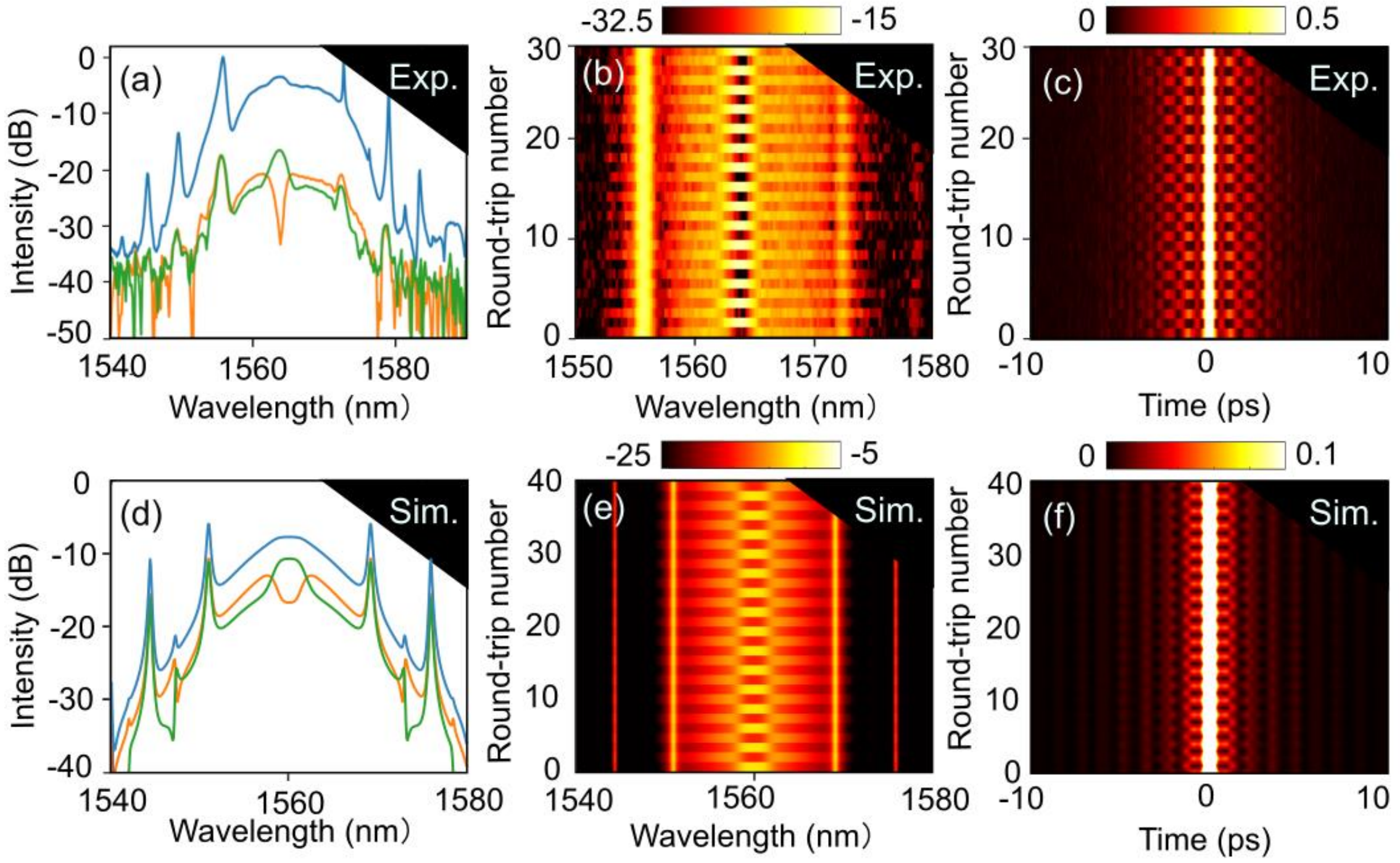

**Figure 1.** Period-2 solitons in an Er-doped anomalous dispersion fiber laser. (a) Time-averaged spectrum (thick blue) and DFT spectra of two successive roundtrips (green and yellow lines). (b) Map of the DFT spectral evolution showing a regular period of 2 roundtrips. (c) Evolution of the first-order temporal autocorrelation trace over roundtrips obtained by Fourier transform of the DFT spectra in Fig. 1(b). (d-f). Simulation results for comparison.

To understand the physical origin of the period doubling, we studied numerically the pulse intracavity dynamics. The pulse spectral evolution is displayed on Fig. 2 (a) and the pulse temporal intensity profile is shown in Fig. 2 (b), along with the intracavity position over two consecutive roundtrips. The corresponding evolutions of the pulse peak power, energy and temporal width are plotted in Fig. 2(c). We note that the pulse energy levels obtained in the two consecutive roundtrips are the same (black curve in Fig. 2(c)), while the pulse peak power (in blue) and pulse duration (in red) evolve in a markedly different way. The invariant change of the pulse energy and the spectral dynamics, dubbed as the “invisible” pulsating soliton behavior,[28] which we found above experimentally for an elementary period-2 pulsation, is here confirmed numerically. The numerical simulation allows us to comment and interpret the dynamics. At the roundtrip N, after exiting from the short normally-dispersive gain fiber, the pulse features a slightly positive chirp (See Fig. 3(a) and (c)). It then propagates into sections of SMF with anomalous dispersion, which results in pulse compression, increasing the peak power until it reaches a maximum at position 31 in the simulation. At this position, the pulse becomes chirp-free with a FWHM duration of 212 fs, see Fig 3(e). Afterwards, the pulse starts to broaden, leading to a decrease in peak power and the change of its chirp to negative, see Fig. 3(g). The spectral peak-to-dip transition occurs along with the flip of the sign of the pulse chirp. Whereas similar temporal breathing dynamics is known to originate from dispersion management,[53] the major spectral breathing and spectral lobes oscillations dynamics indicate a prevalence of the self-phase modulation.[54] Such prevalence means an excess of the nonlinearity over the averaged anomalous dispersion, which destabilizes the period-1 regime. Indeed, during the following roundtrip (N+1), we can understand the spectral dip-to-peak dynamics as a healing of the excess of nonlinearity experienced during the roundtrip N. The pulse peak power during the roundtrip N+1 remains significantly lower than during the roundtrip N, which also results from the prevalence of the positive frequency chirp, and a minimum pulse duration of 260 fs. As the above dynamics mainly originates from an excess of the Kerr nonlinearity, one could see a connection with the dynamics of high-order bright solitons of the NLS equation with anomalous dispersion. However, this connection is quite distant, as the breathing pulse remains a single bell-shaped profile, in contrast to the periodic pulse breaking experienced for high-order conservative solitons. Therefore, whereas dissipative soliton dynamics in anomalous average dispersion regime can take some attributes of NLS soliton dynamics, it cannot be reduced to it.

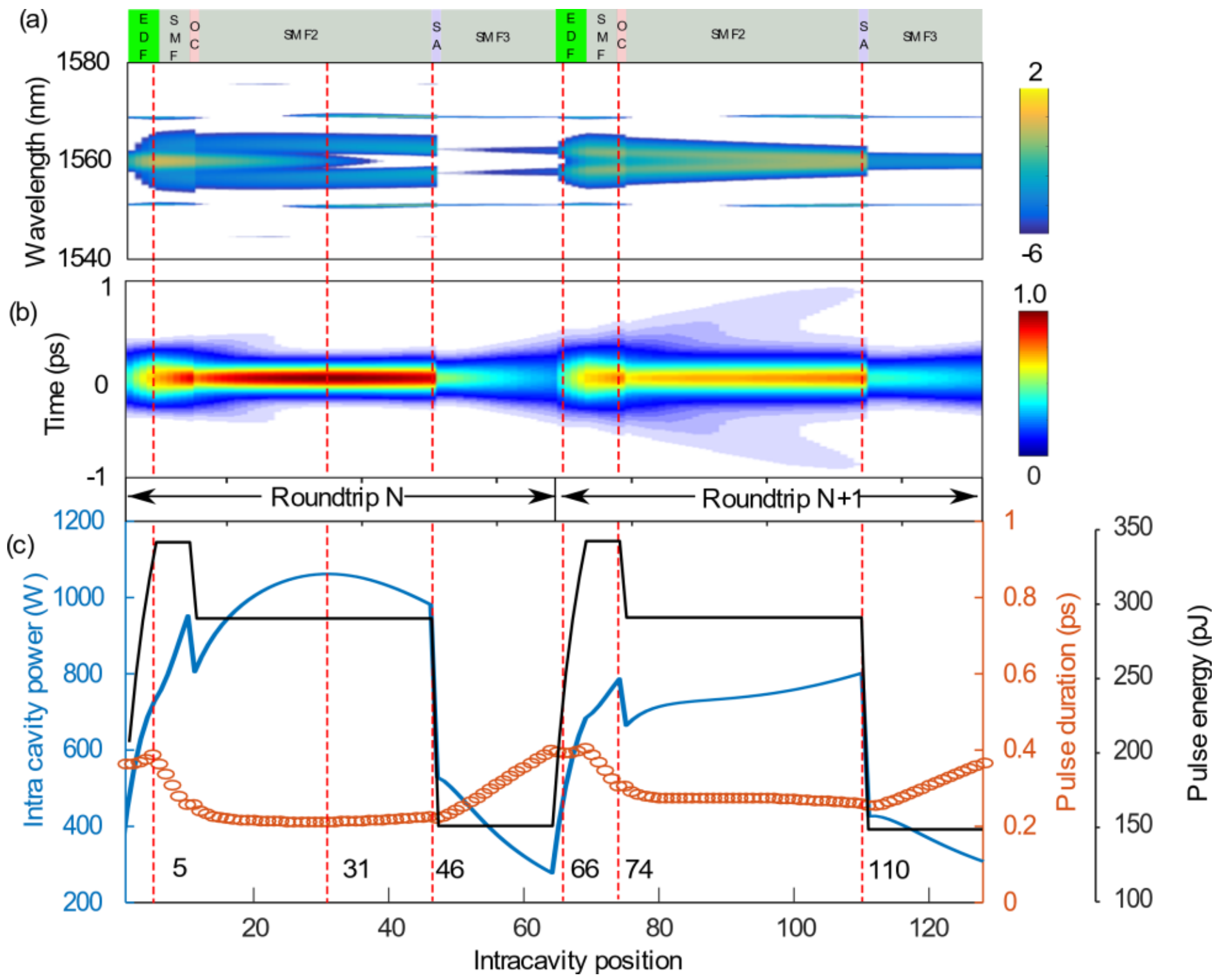

**Figure 2.** Simulation of the intracavity light field evolution in the case of the period-2 pulsation, over 2 consecutive roundtrips. Roundtrip N is from slice 1 to 64. Roundtrip N+1 is from slice 65 to 128. (a) Spectral intensity profile evolution. (b) Pulse intensity profile evolution. (c) Evolution of the pulse energy (black curve), peak power (blue curve) and pulse duration (orange curve).

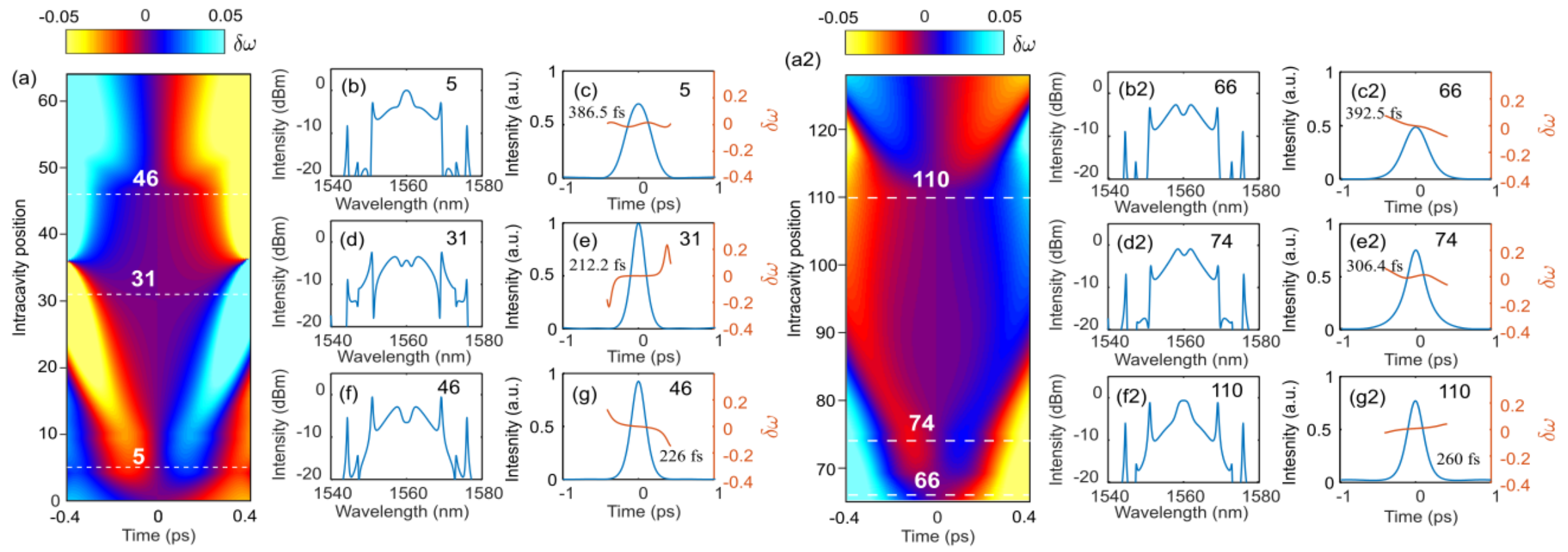

**Figure 3.** Simulation of the period-2 pulsation, showing the evolution of the spectrum and the chirp of the central part of the pulse during the roundtrip N (left columns) and N+1 (right columns), highlighting the peak-to-dip and dip-to-peak spectral transition dynamics, respectively, associated with the change of the sign of pulse chirp.

## 2.2. Toward instability: long-period and multiple-period pulsations

For a slightly higher pump power, in the 80-90 mW range, the periodicity of the spectrum oscillations can evolve in an intriguing fashion. The DFT evolution map in Fig. 4 (a) shows oscillating spectral patterns with the central part of the optical spectrum changing periodically, with a recurrence happening every 7 roundtrips. This period-7 unusual periodicity is better seen in Fig. 4 (b), where the intensity close to the wavelength $\lambda$ = 1571 nm (in black), extracted from the DFT measurement, is plotted against the number of cavity roundtrips. The total intensity of the pulse (in red) is also plotted, showing that it remains nearly constant within the noise fluctuations of the oscilloscope, making another illustration of a so-called “invisible” pulsation. We also indicate that the time-averaged optical spectrum measured by an OSA looks like those recorded for the period-2 oscillations cases. The period-7 pulsation revealed here cannot originate from a mere sequence of period-2 bifurcations: another type of bifurcation is at stake. Nevertheless, the alternation of spectral values – except for two consecutive ones – indicates a likely interplay of period doubling with other different bifurcations.

Higher periodicities can also be obtained. In the example shown in Fig. 4 (d, e, f), the DFT evolution map displays a blend of complex and stable patterns, depending on the spectral window analyzed. The intensity of the central part of the spectrum at 1565 nm is periodic with a period larger than 10. A Fourier transform of the latter is performed on a full acquisition (1600 roundtrips) to obtain a precise evaluation of the periodicity of the soliton spectrum. The spectral density of Fig. 4 (f) reveals that the lowest periodicity of the soliton spectrum is not an integer, with a value of T = 11.6 roundtrips. A larger, broader peak is also obtained at a frequency of 0.5, corresponding to the period-2 contribution to the oscillations, clearly manifested in the alternation of spectral intensities in Fig. 4(e). From this working point, by slightly increasing the pumping power, we reach a destabilization of the oscillating patterns. Fig. 4 (g, h, i) show the case where a high-order periodicity is still present, but the oscillation amplitude varies in time. As a result, the spectral content of the intensity of the oscillating part of the spectrum displays a broadband peak, centered around a period T $\sim$ 14.5 roundtrips. With numerical simulations, by using a gain parameter $g_0$ = 16.7 m$^{-1}$, we obtained qualitatively similar results, as shown in Fig. 4(j, k, f) with period-2 and period-32 oscillations.

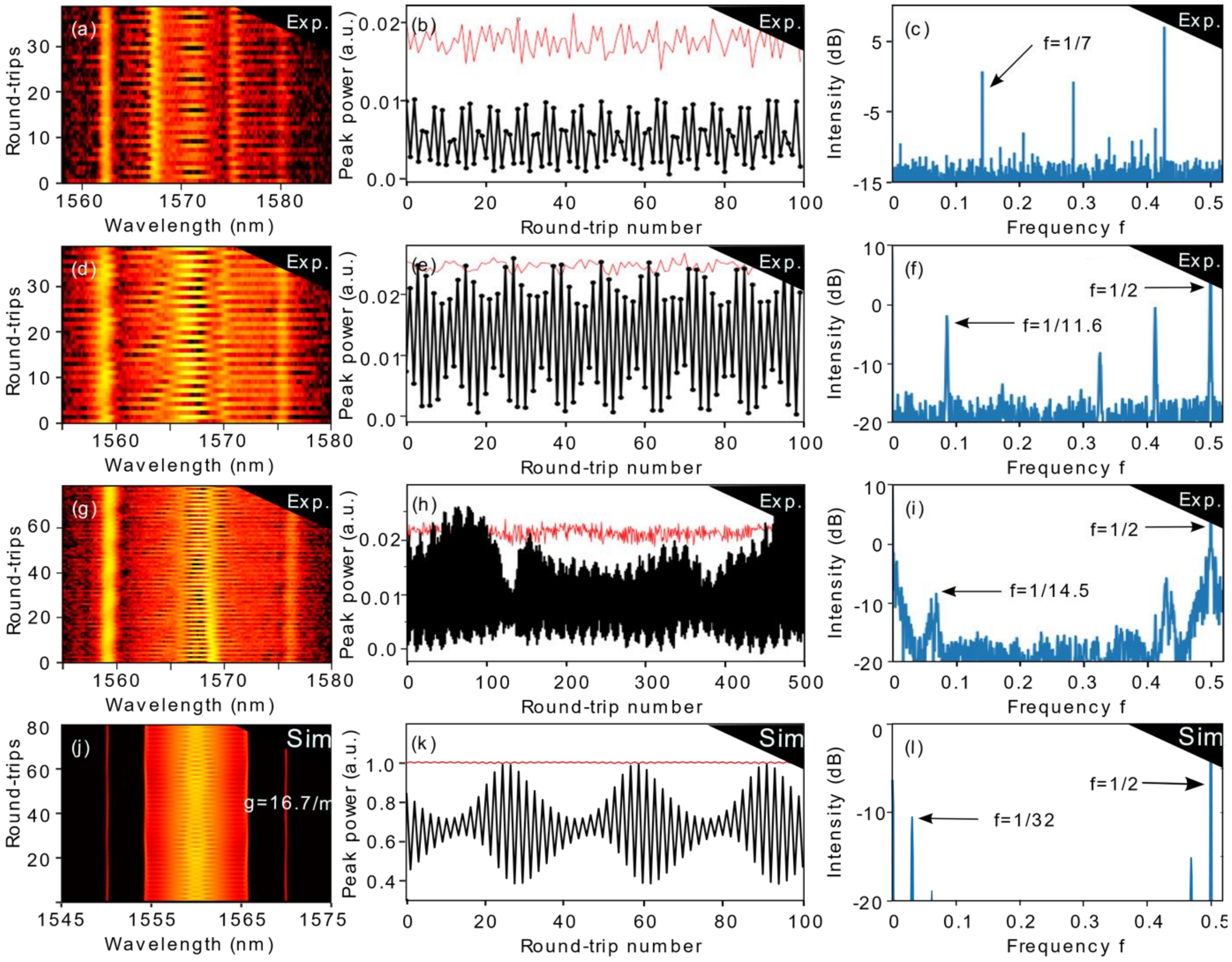

**Figure 4.** Characterization of 3 different regimes where period multiplication and instabilities occur. (a, d, g) are evolution maps of the DFT measurements, (b, e, h) plot the evolution of the pulse intensity (red) and the oscillating region of the spectrum (black) as a function of the round-trip number, and (c, f, i) show the Fourier transform of the latter, so that the periodicity appears clearly. The first regime has a 7-roundtrip periodicity, the second is close to 12-period and the third has a broad periodic contribution around 14.5 roundtrips. (j-l) simulation results at a small signal gain value of 16.7 $m^{-1}$, showing a qualitative match with the experimental observations.

### 2.3. Observation of frequency entrainments

To understand better the bifurcation cascades involved in the previous laser dynamics, we recorded the transient dynamics of the bifurcation sequences starting from a single-period soliton and leading to period-N solitons, by ramping up linearly the pump power from 70 mW to 90 mW. The oscilloscope is triggered on a higher level of the DFT signal. Figure 5 (a) represents the DFT evolution map over more than 15000 cavity roundtrips as the pump power is increased and reveals that the oscillations quickly appear. To shed more light on the phenomenon and timescale at stakes, the evolution of the intensity in the oscillating part of the spectrum at $\lambda_i$ ,the center wavelength, is plotted as a function of the roundtrip number in Fig. 5 (b). The first bifurcation after 4000 roundtrips leading to the 2-period pulsation is clearly visible, followed by a second bifurcation at 8000 roundtrips, which this time corresponds to the beginning of the modulation of the envelope of the oscillations. Between 10500 and 12500 roundtrips, a structure emerges in the oscillation pattern, revealing the organization of the oscillating patterns into an integer periodicity. For a better comprehension, one out of every three data points of the same acquisition has been plotted in red, with 4 branches appearing clearly, hence a 12-periodicity. To follow more accurately the evolution of this modulation, a spectrogram of the intensity at $\lambda_i$ is shown on Fig. 5 (c): for each round-trip, we calculate the Fourier transform of the next 1024 points and plot the evolution of the spectral density (in log scale) as the round-trip number increases. In the first 4000 roundtrips, the spectral density is flat, indicating that the intensity at $\lambda_i$ is not oscillating. After the first bifurcation, a signal at a frequency f = 0.5 appears rapidly, corresponding to the 2-period oscillations. The second bifurcation is also clearly visible at 8000 roundtrips, with the apparition of a signal at a frequency f $\sim$ 0.08. The spectrogram reveals that this frequency continuously drifts towards higher value, until it reaches f = 1/12 where it remains constant for 2000 roundtrips. This regime corresponds to the organization seen on Fig. 5 (b), with a 12-periodicity. After 12500 roundtrips, the peak frequency starts again drifting towards higher frequencies. The second line of Fig. 5 shows a similar study with a slightly different initial state. The pump power ramp is identical to the previous case, and two successive bifurcations also appear in Fig. 5 (e), though in this case, the final state corresponds to a 14-periodicity, as shown by the 1 out of every 7 data points plotted in red. The spectrogram of Fig. 5(f) confirms this analysis, as well as the continuous drift in frequency of the lowest frequency peak, finally settling at f = 1/14.

Simulation results of the spectrum transformation from stable solitons to period-N solitons are shown in Fig. 5(h-j). The small signal gain is increased from 14 $m^{-1}$ to 16.4 $m^{-1}$ by 0.1 $m^{-1}$ steps and from 16.4 $m^{-1}$ to 16.8 $m^{-1}$ by 0.01 $m^{-1}$ steps. For each gain value, we run 1000 roundtrips. The initial condition is a white noise field and the final output light field at the current value of gain is used as the initial condition for the next. The spectral evolution shows transitions from a stable period-1 mode-locking of solitons at 14 $m^{-1}$ to a multiple periodicity at 16.8 m$^{-1}$. Fig. 5(i) depicts the pulse energy (grey dots) and the intensity of the central spectrum at $\lambda_i$ (red dots) as a function of gain. A clear period doubling bifurcation and a quasi-period-N bifurcation appear, whereas the pulse energy slowly increases without revealing these bifurcations. A spectrogram of the intensity of the central spectrum as a function of $\lambda_i$ is calculated and shown in Fig. 5(k), revealing the birth of the period doubling at $g_0$ = 14.8 m$^{-1}$ and the occurrence of the new frequency f ≈ 1/32 at $g_0$= 16.66 m$^{-1}$. This frequency is slightly red shifted as $g_0$ increases from 16.66 m$^{-1}$ and finally stabilized at f = 1/32. Therefore, the simulation results qualitatively reproduce the experimental observations. Following the bifurcations leading to LPP, we observe an entrainment of the LPP frequency and its locking to an integral number of roundtrip numbers. Our experimental and numerical observations are consistent with recent similar reports of the entrainment of LPP – or breathing solitons – in driven Kerr microresonators and ultrafast lasers.[55,56]

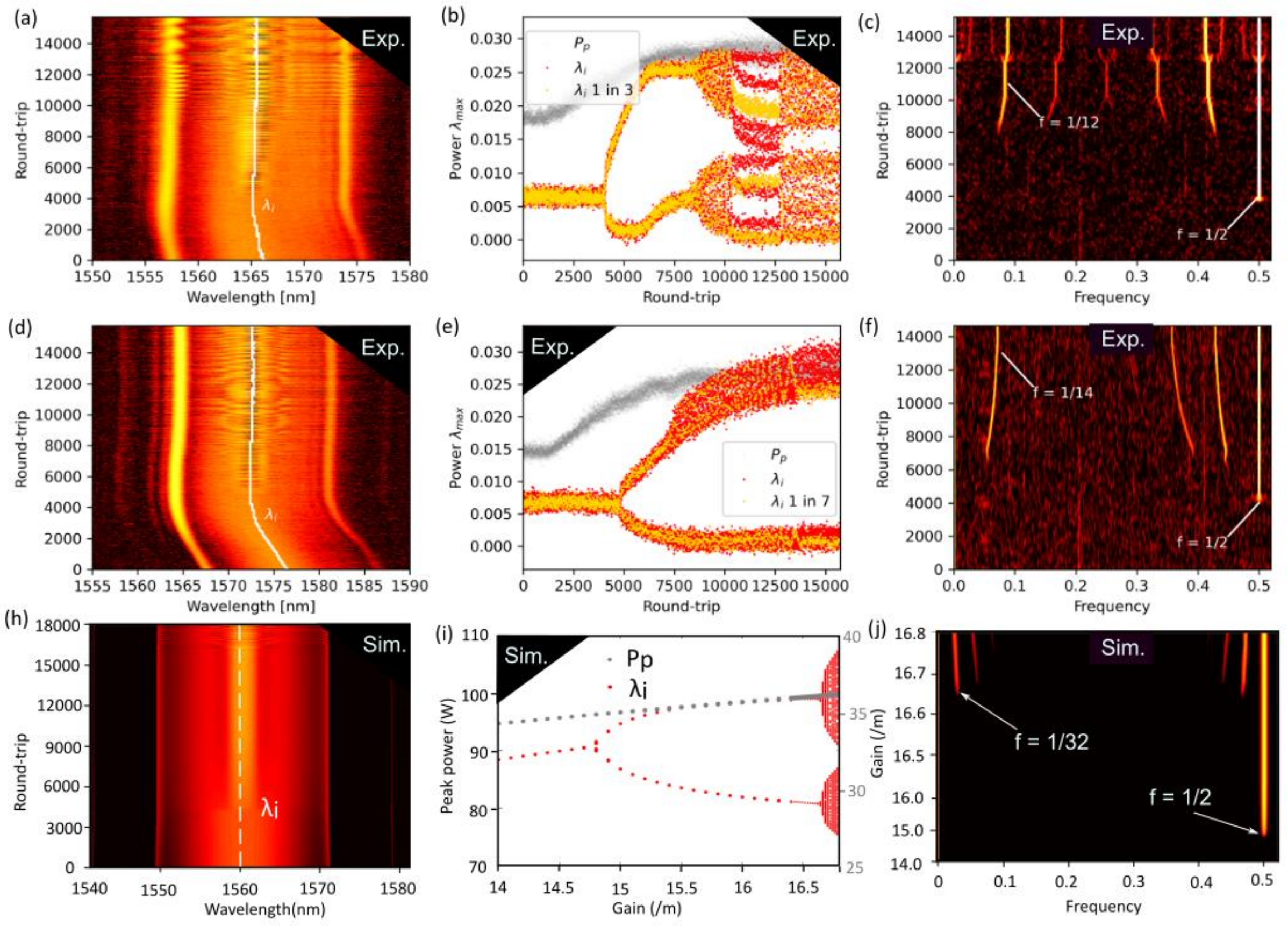

**Figure 5.** Transient dynamics of pulsating solitons under a millisecond ramp up of the pump power. Two experimental recordings having slightly different initial parameters comprise: (a, d) The consecutive DFT output spectra, where the wavelength of reference $\lambda_i$ is indicated. (b,

e) The evolution of the spectral intensity at $\lambda_i$, showing the sequence of period-2 and period-N bifurcations. (c, f) The Fourier analysis of the spectral intensity pulsation at $\lambda_i$. A numerical simulation is presented in (h-j) for qualitative comparison, using an increase of the gain parameter from 14 $m^{-1}$ to 16.8 $m^{-1}$.

## 3. Period doubling in the normal dispersion regime

In the anomalous dispersion regime (Section 2.1), we discussed a possible – though distant – connection between the period-2 pulsation of the dissipative solitons and the dynamics of high-order bright solitons of the NLS equation. Now, considering the propagation of bright dissipative solitons in a normal dispersion regime, the prospect of such comparison seems vanishing. However, we first need to check whether the period-doubling bifurcation will take place to further confirm its universality in the dynamics of dissipative solitons.

We built a normally-dispersive Er-doped mode-locked fiber laser whose experimental details are provided in the Experimental Section of the Supplementary Materials. At a pump power of 400 mW, the laser yields a stable mode locking (see Fig. 6 (a-c)). The real-time recording of consecutive single-shot DFT spectra confirms the stable mode locking operation, see Fig. 6(a) and (b). The structured optical spectrum with sharp edges is typical of the normal dispersion regime.[57,58] The radio frequency (RF) spectrum shown in Fig. 6 (c) indicates a single peak located at 20.1 MHz that corresponds to the fundamental pulse repetition rate and evidences the mode locking stability.

As a matter of fact, period-2 pulsating solitons are observed through solely increasing the pumping power, as in the anomalous case described in the previous sections. Figure 6 (d) shows the evolution of the DFT-recorded spectrum over 20 roundtrips at a pump power of 437 mW. There exist two distinct spectral shapes in two consecutive roundtrips and there is an abrupt transition between them. The spectrum shown in Fig. 6(e) reflects an energy flow from the center part of the spectrum in the $N^{th}$ roundtrip to the edges of the spectrum in the $N+1^{th}$ roundtrip. The evolution of the pulse energy (white line in Fig. 6(d)) clearly reflects the period doubling regime: this time, we do not have an "invisible pulsation". The RF in the period-doubling regime exhibits two symmetric sub-sidebands around the central peak, separated by 10.05 MHz, which is half of the fundamental pulse repetition rate. Further increasing the pumping power leads to a multi-pulse operation or disrupts the mode-locking operation.

Therefore, the increase of the pumping power has here three main outcomes: period doubling, multi-pulse operation or chaotic behavior.

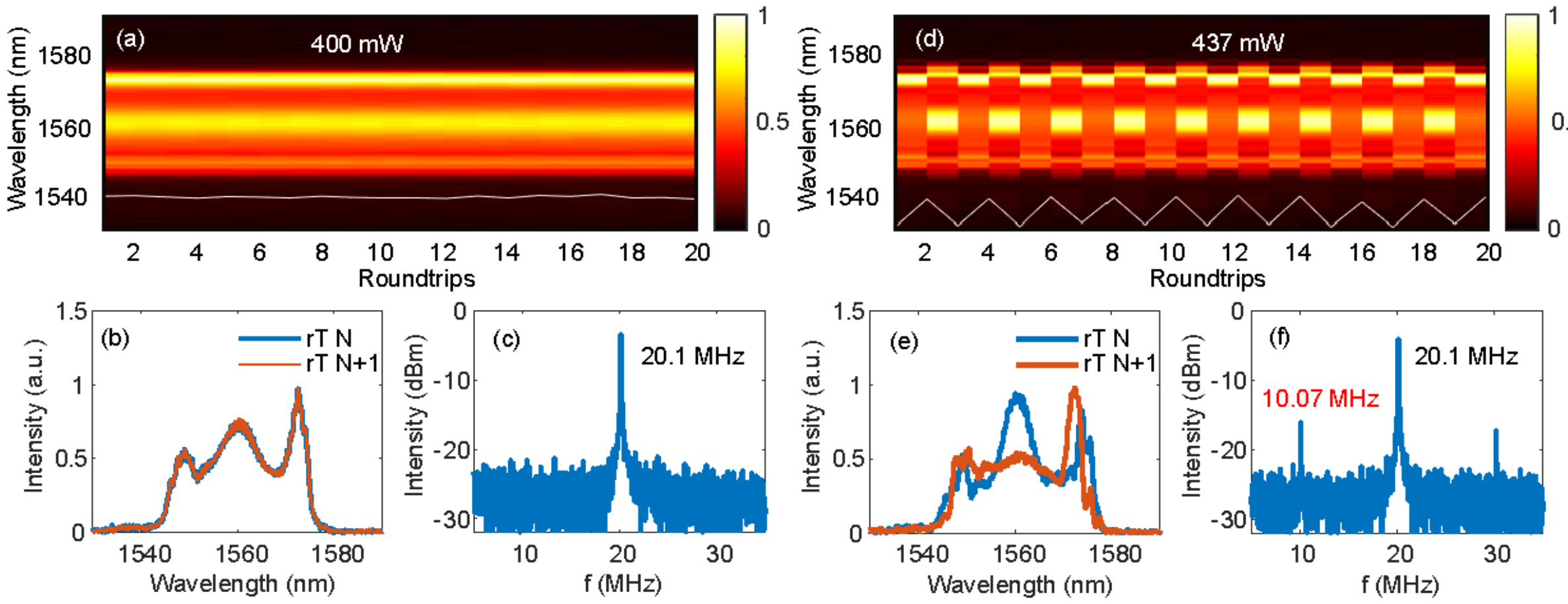

**Figure 6.** Experimental results of period-2 bifurcation in the normal dispersion regime. (a, b, c): Stable mode locking operation at 400 mW. (a) Spectrum evolution. The white curve is the evolution of the pulse energy. (b) Two consecutive single-shot spectra and (c) the RF spectrum. (d, e, f): Period doubling dynamics at 437 mW. (d) Spectrum evolution. (e) Two consecutive single-shot spectra and (f) RF spectrum. 20.1 MHz corresponds to the fundamental cavity repetition rate.

We carried out numerical simulations to corroborate the experimental observations and interpret the intracavity dynamics. The threshold for stable mode locking is obtained at a gain parameter $g_0$=3.5 m$^{-1}$. Figure 7(a) displays the succession of 10 output optical spectra in a stable mode-locking regime with $g_0$ =8 m$^{-1}$. The spectrum is centered at 1560 nm with sharp edges on both sides. As expected from the normal dispersion regime, the output pulse features a positive chirp (red dashed curve in Fig. 7(c)). Further increasing the gain beyond 11.5 m$^{-1}$ leads to the period doubling bifurcation. A sequence of the bifurcation versus gain is shown in Fig. S4 in the Supplementary material, revealing the bifurcation of the pulse energy and peak power in the normal dispersion regime. Different from the "invisible" pulsations in the anomalous dispersion regime,[28] the simulated period-2 pulsation confirm the visible pulsations observed in the dissipative normal-dispersion regime. Figure 7(d) shows successive output optical spectra for $g_0$=12 m$^{-1}$. Two consecutive laser outputs pulses largely differ in their spectral shape and energy (Figs. 7 (d, e)). The pulse peak power in the period doubling regime exceeds by far that of the stable mode locking, while the pulse duration is narrowed down to the 0.1-ps level, to be compared with the picosecond pulse duration obtained in the stable mode locking regime. Such major differences indicate the excess of nonlinearity in the period doubling regime.

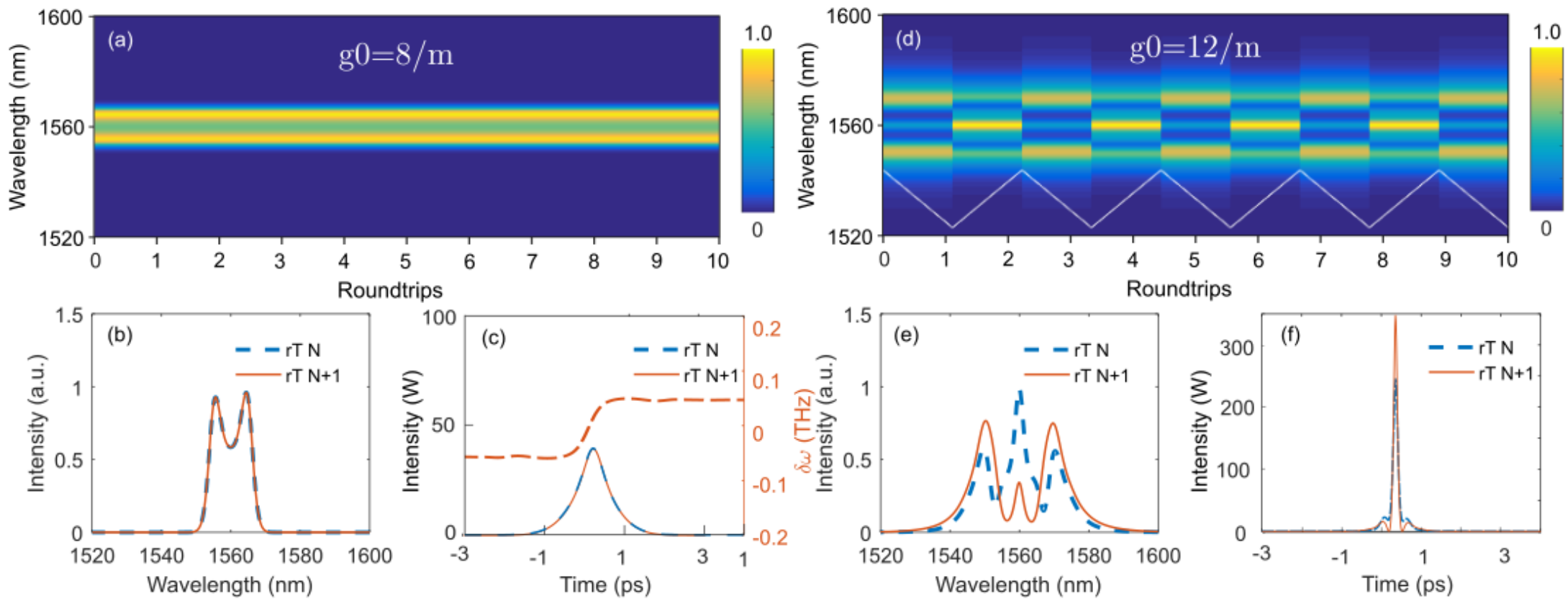

**Figure 7.** Simulation results of the dynamics of the fiber laser in the normal dispersion regime. (a-c): Stable mode locking operation. (a) Optical spectra over successive roundtrips. (b) Two consecutive optical spectra (c) Output pulse profile and pulse chirp. (d-f): Period doubling operation. (d) Optical spectra and pulse energy (white curve, with offset) over successive roundtrips. (e) Two consecutive optical spectra and (f) Output pulse intensity profile.

By displaying the evolution of the intracavity fields, we obtain more insight into the period doubling mechanism. The intracavity evolution in the stable mode-locking operation regime is shown in Fig. 8(a-c). The pulse evolves self-consistently over successive roundtrips. In the normally-dispersive gain fiber (EDF), the pulse is broadened and amplified. Then it is compressed in the anomalous SMF1 segment. Due to the nonlinear self-phase modulation (SPM) effect, new frequencies emerge on the spectrum during the pulse compression. An optical coupler extracts 10% of the intracavity energy and the remaining 90% passes through a SA and a Gaussian filter. The 6-nm bandwidth Gaussian filter narrows significantly the spectrum and decreases the pulse energy and peak power, as shown in Fig. 8(c). The pulse is broadened in the dispersion compensation fiber (DCF) endowed with a large normal dispersion. The pulse output from the DCF has the same characteristics as the initial input pulse for the next roundtrip. Therefore, the pulse in the stable mode-locking regime repeats its evolution every roundtrip. We note that whereas the laser cavity includes dispersion management, i.e. propagation within a succession of normal and anomalous dispersion, the propagation regime is here typical of a net normal dispersion regime, with the pulse duration having one local minimum per roundtrip, in contrast to dispersion-managed soliton propagation with two minima per roundtrip.[59]

In the period doubling regime, the main difference takes place during the pulse compression process in the SMF1. The pulse, after gaining a larger energy through the EDF, experiences a strong SPM effect and is over compressed, which is reflected by the non-monotonic evolution of the pulse duration (see red curve in Fig. 8(f)). The wave-breaking-like structured spectrum in Fig. 8 (d) and the compressed pulse with large pedestal component in Fig. 7(f) confirm the strong nonlinear compression stage. In the $(N+1)^{\text{th}}$ roundtrip, the pulse exiting the EDF has a longer duration (2.5 ps) and a larger positive chirp than after the EDF in the $N^{\text{th}}$ roundtrip. In the $(N+1)^{\text{th}}$ roundtrip, the pulse propagating in SMF1 is nearly monotonically compressed to a chirp-free pulse (See in Fig. 9(f)). After SMF1, the combined effects provided by OC, SA and the spectral filter reshape the pulse profile and ensure the self-consistent evolution. Therefore, the excess of Kerr nonlinearity plays a major role in the period doubling in the normal dispersion regime as well. The comparison between period-1 and period-2 regimes highlights how a slight change in the pulse parameters (duration, chirp and peak power), pushes the laser toward different dynamics.

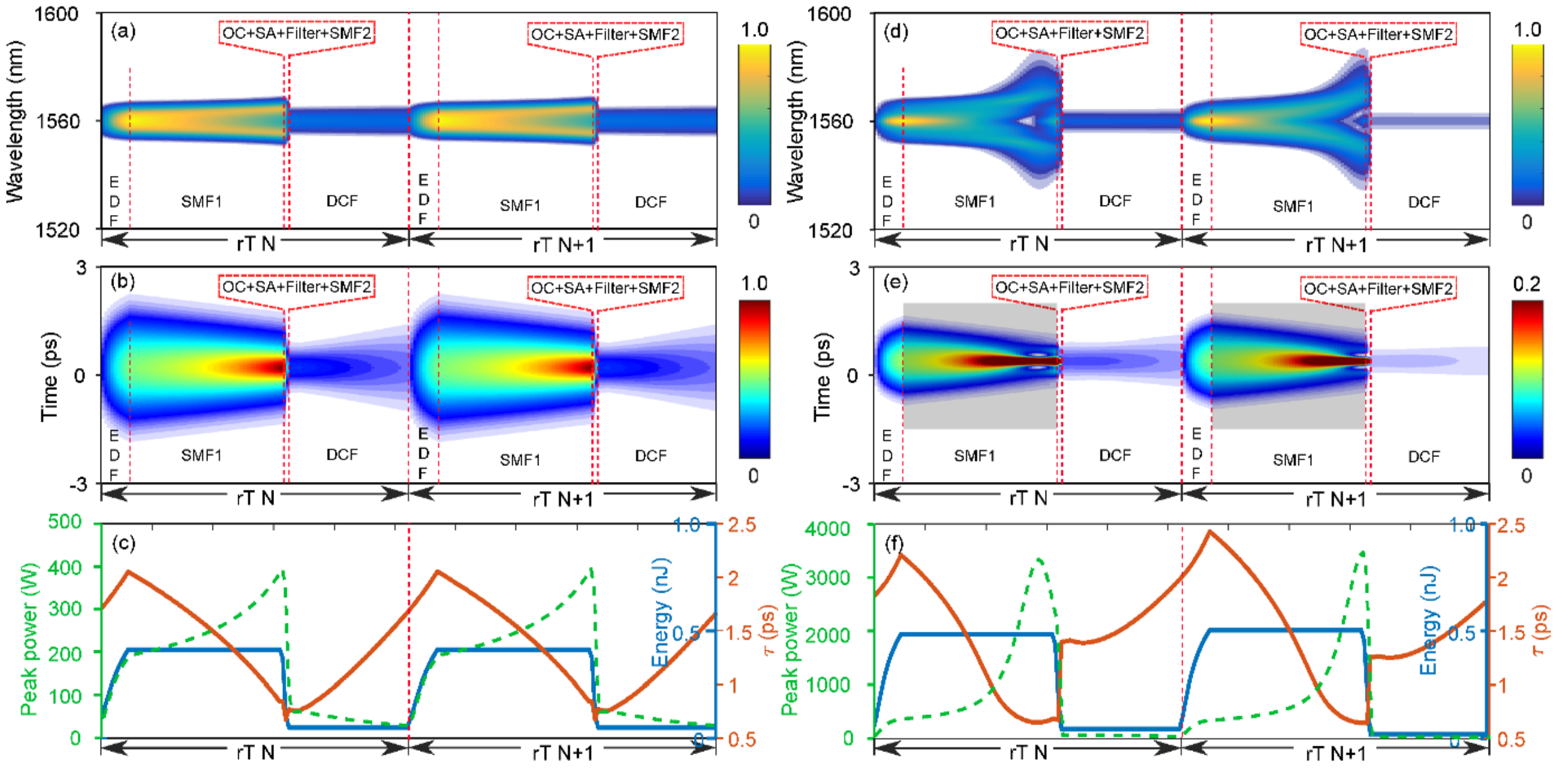

**Figure 8.** Simulation results of the intracavity pulse dynamics over two consecutive roundtrips. (a-c): Stable mode-locking regime. Evolution of (a) the optical spectrum, (b) the temporal

intensity profile, (c) the peak power (dashed green line), pulse energy (cyan line) and pulse duration (orange line). (d-f): Similar monitoring for the period-2 regime.

For the period-2 pulsation, we present in Fig. 9 a zoom-in of the simulated pulse evolution in the SMF1, marked as in grey boxes in Fig. 8 (e), as well as the corresponding pulse chirp evolution. The false color plots of the evolution of the pulse chirp of the central part of pulses in Fig. 9(c) and (d) evidence the major role of the nonlinear pulse compression in the period doubling regime. In the $N^{\mathrm{th}}$ roundtrip, the initially positive pulse chirp becomes negative at the end of the SMF1. In the $(N+1)^{\mathrm{th}}$ roundtrip, the input pulse has a slightly lower peak power and longer duration (See in Fig. 9(g)), thus experiencing a weaker SPM effect in the compression process. Consequently at the $(N+1)^{\mathrm{th}}$ roundtrip, the pulse exiting from the SMF1 becomes nearly Fourier-transform-limited and displays significant pedestals owing to the overall importance of the spectral broadening through SPM that is obvious from Fig. 8(d).

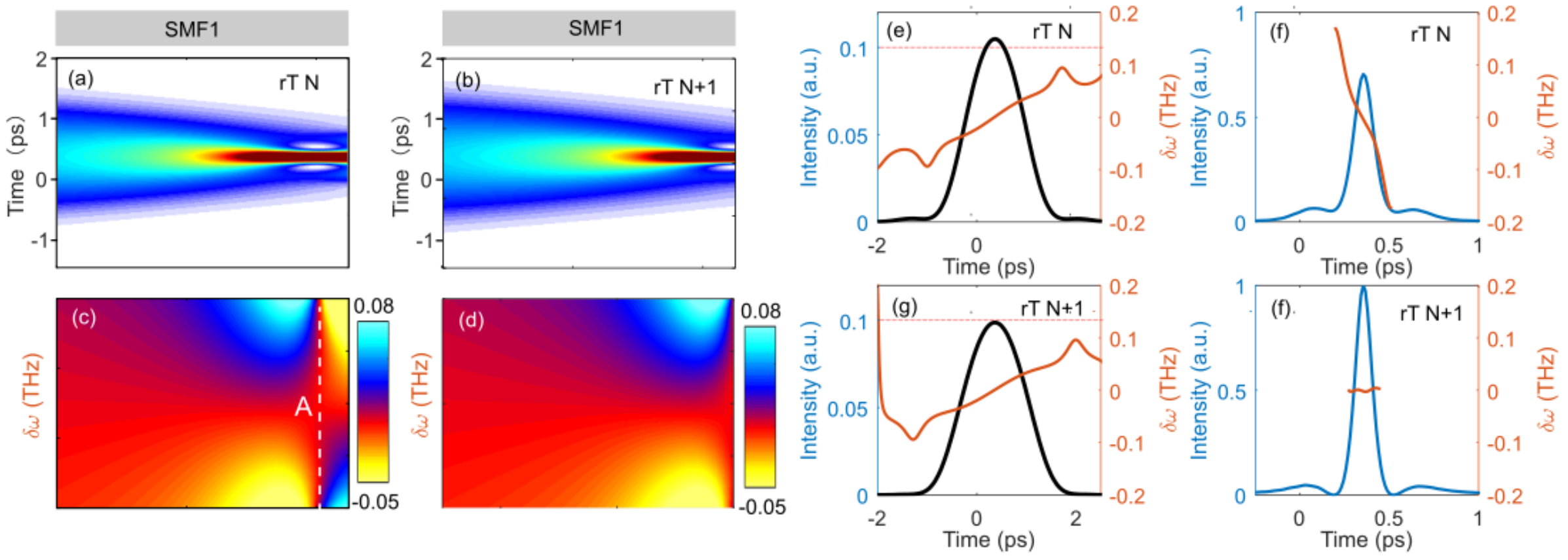

**Figure 9.** Dynamics of the nonlinear pulse compression process taking place in SMF1 in two consecutive roundtrips. (a) and (b): Temporal pulse profile; (c) and (d): Chirp profile. The color represents the value of the instantaneous frequency. In (c), the dotted line A indicates the chirp-free point. (e) and (f) depict the pulse profile before and after the SMF1 in the $N^{\mathrm{th}}$ roundtrip. (g) and (h) depict the pulse profile before and after the SMF1 in the $(N+1)^{\mathrm{th}}$ roundtrip.

## 4. Discussion and Conclusion

After presenting an extensive introduction to the field of pulsating dissipative optical solitons, we have provided an original research work based on several fiber laser experiments, thus complementing previous observations. We observed and simulated period-2 pulsations in ultrafast fiber lasers under different dispersion regimes, confirming the universality of this bifurcation within ultrafast lasers. The period-2 dynamics manifest as peak-to-dip (dip-to-peak) transformation of the central intensity of the spectrum in the frequency domain. Our numerical modeling indicates that an excess of the Kerr nonlinearity has a leading role in the manifestation of the period-2 bifurcation, which is why it often constitutes the aftermath of an increase of the pumping power, starting from a stationary period-1 mode locking regime. We have also highlighted how the period-2 bifurcation can be followed by period-N (N>2) bifurcations, leading to doubly periodic pulsations, which can be viewed as one category of LPP arising from the overdrive of the pumping power. These LPP are also prone to destabilization, chaos, or complete disruption of the short-pulsed regime whenever the pumping power is subsequently increased. This way, the appearance of the period-2 pulsation can serve as an early warning signal before the disruption of the short-pulsed regime. However, we noted that, despite a common major role of the SPM effect in both anomalous and normal dispersion regimes, the period-2 pulsation manifests differently through the optical characterization devices. In the anomalous regime, our recorded period-2 pulsations were nearly "invisible", to use the qualifier introduced in [28], which means they did not entail a noticeable pulse energy change over consecutive cavity roundtrips. In contrast, in the normal dispersion case, the period-2 pulsations involved a significant alternation of the output pulse energy. Therefore, a reliable early warning signal could be built from the shot-to-shot spectral recording, for instance within a given spectral slice.

There is another category of LPP bifurcation, which can directly lead to LPP regimes without going through the period doubling bifurcation.[16] In ultrafast lasers, such direct LPP bifurcation appears when the energy of the nonlinear system is decreased, which typically happens when the pump power is lowered below the stable mode-locking threshold.[25,27] Therefore, reaching the direct LPP bifurcation involves playing around with the hysteresis that surrounds mode locking. In driven Kerr cavities, LPP can also involve the manipulation of hysteresis: when the driven power is fixed and the detuning, namely the frequency offset between the driven laser and the cavity resonance frequency, sweeps the resonance from the blue detuned to the red detuned region, LPP are observed before the onset of a stable Kerr cavity soliton operation, at lower pulse energy.[33,60,61] In the nonlinear dynamics of driven cavities, such LPPs have been called "breathers" and the direct LPP bifurcation can also be accessed from a stable cavity soliton regime upon the increase of the pump power, highlighting differences with ultrafast laser dynamics.[62] By extension, the notion of LPP applies to complex pulse patterns such as oscillating optical soliton molecules, where the pulsation involves the internal degrees of freedom of the soliton molecule, namely the relative separation and phase between adjacent pulses.[19] But it is also possible to observe pulsations of soliton molecules that do not involve their internal degrees of freedom, see the example provided in §S1.2 of the Supplementary Materials.

We have also shown experimentally and confirmed numerically the phenomenon of synchronization of the LPP periodicity with integer multiples of the cavity roundtrip, within a finite range of the laser parameters. The frequencies of oscillations entrained in the sequence of bifurcations from stable solitons to quasi-periodic N solitons tend to lock to an integral number of roundtrips. In our experiments, this frequency locking is observed at a pump power beyond the threshold of stable mode-locking. In the supplementary material section S1.3, we provide an additional example obtained in the wavelength region around 1.9 μm, by using a thulium-doped fiber laser. Whereas the synchronization of periodic pulsations was predicted quite a few years ago,[16] our experimental observations confirm the universal reach of synchronization phenomena, also illustrated recently in the case of the subharmonic entrainment of LPP breather solitons in both driven Kerr microresonators and ultrafast lasers.[55,56] The coexistence of the period-2 frequency with the entrained frequency of quasiperiodic-N, which is synchronized to the roundtrip time, define multifrequency clock speeds, which could have a great applied potential in optical communications.

## 5. Experimental Section

*Anomalous dispersion ultrafast laser*: The anomalous dispersion fiber laser used in our experiment consists in a ring cavity shown in Fig. 10. A 50-cm long erbium-doped fiber (EDF, 110 dB/m absorption, 4 μm core diameter) provides gain to the fiber cavity and is pumped by a 980-nm, 900-mW pump diode. The multiplexer used in the cavity also acts as an isolator for the signal at 1550 nm. A fiber-based polarization controller and polarization beam splitter complete the fiber cavity and provide an effective saturable absorber function based on the nonlinear polarization evolution (NPE) that takes place in the intracavity fibers. By adjusting the position of the polarization controller, one can explore a wide variety of effective saturable absorber transfer functions and resulting laser dynamics. A careful adjustment of the polarization controller allows for mode-locked operation with a single soliton for a pump power as low as 70 mW. The cavity roundtrip time is 31.7 ns (repetition rate of 31.5 MHz) corresponding to an overall length of 6.3 m split in 5.8 m of standard single-mode fiber (SMF) and 50 cm of EDF. The group velocity dispersion at 1.55 µm is 13.5 $ps^2\ km^{-1}$ for the EDF and $-22.9\ ps^2\ km^{-1}$ for the SMF, yielding an anomalous averaged cavity dispersion parameter $\beta_2 = -20\ ps^2\ km^{-1}$ and a total dispersion of $-0.126\ ps^2$. The slow-axis output of the polarization beam splitter constitutes the output of the laser and is sent to the various instruments used for characterization.

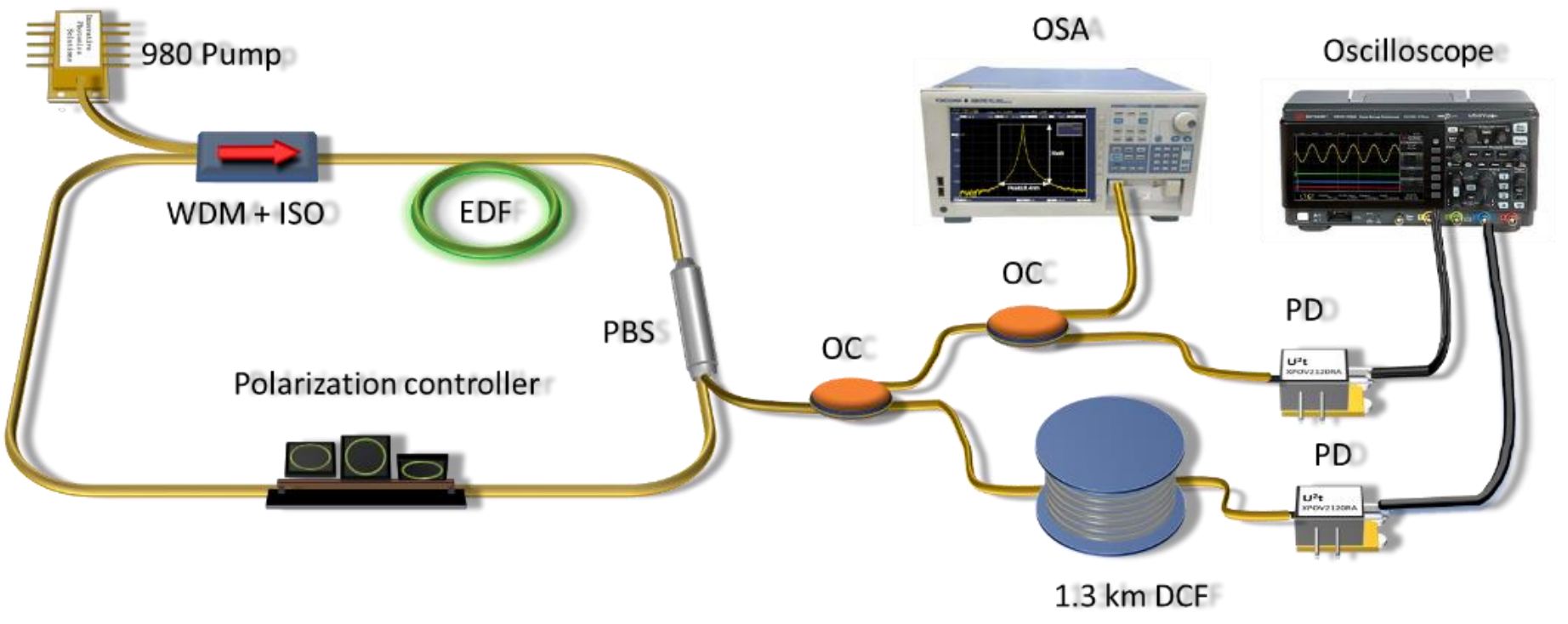

**Figure 10** Scheme of the experimental setup involving the anomalous dispersion fiber laser. The laser consists in a 50-cm erbium-doped fiber (EDF) pumped by a 980 nm-laser diode through an isolator-multiplexer, a single-mode fiber (SMF) polarization controller and a polarization beam-splitter (PBS). The slow-axis output of the PBS is split and sent to the analysis instruments: a time averaged optical spectrum analyzer (OSA), a fast photodiode, and a 1.3 km-long dispersion compensation fiber (DCF) to perform the dispersive Fourier transform (DFT). An autocorrelator was also used to measure the auto-correlation trace of Fig. 3 (b).

*Normal dispersion ultrafast laser*: The normal dispersion ultrafast laser we used to study the period doubling is an EDF ring laser mode locked via the NPE technique. The laser consists of a 1-m highly doped EDF, 3.6 m dispersion compensation fiber (DCF) and 5.35 m of single mode fiber (SMF), respectively. The group velocity dispersion of the EDF, SMF, and DCF are 61.2 $ps^2/km$, -22 $ps^2/km$ and 56.1 $ps^2/km$ at 1550 nm, respectively. Consequently, the laser yields a net normal dispersion of 0.145 ps$^2$. A 980-nm laser diode with a maximum power of 700 mW is used to forward pump the EDF and a polarization-dependent isolator (PD-ISO) is inserted in the cavity to suppress the backward stimulated Brillouin scattering and ensures the unidirectional lasing operation. A combination of two polarization controllers and the PD-ISO allow to adjust the artificial SA to mode lock and stabilize the pulses. 10% of the internal cavity power is extracted out from a 10:90 optical coupler for the measurement. The total cavity length is 9.95 m, resulting in a roundtrip time of 49.8 ns and a fundamental repetition rate of 20.1 MHz.

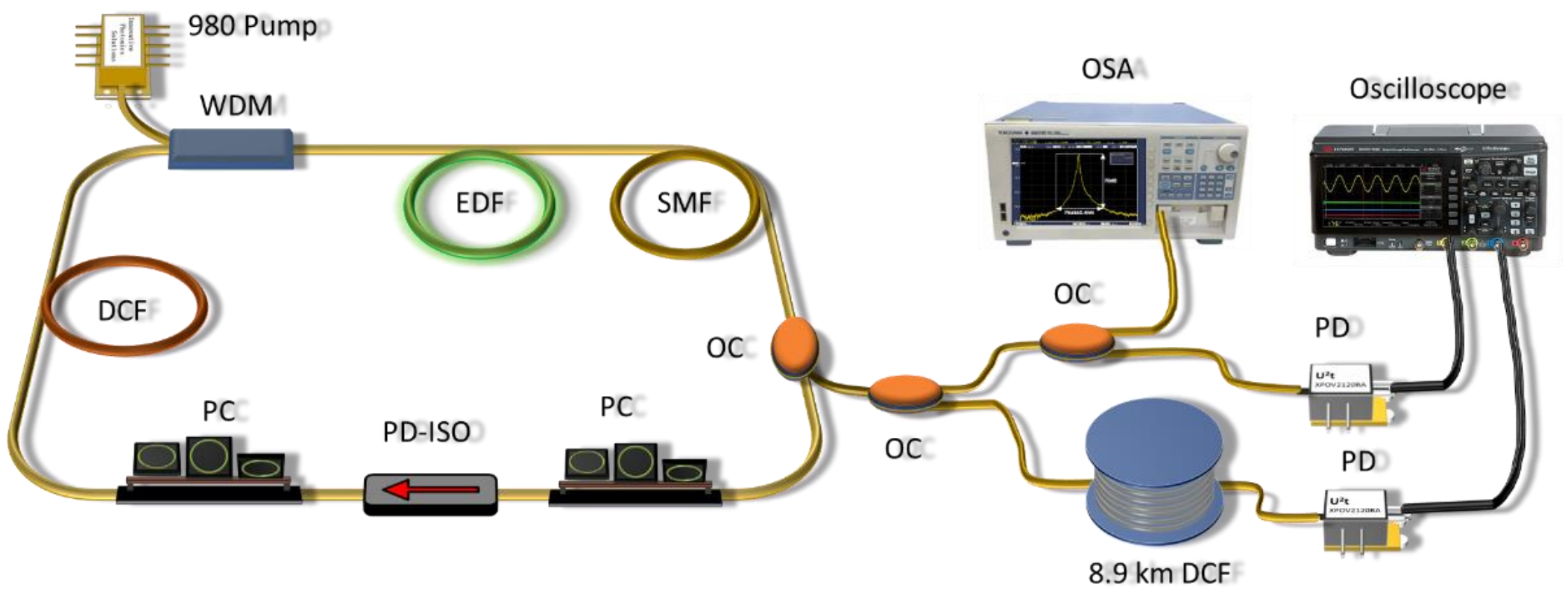

**Figure 11.** Experimental setup for observing the period doubling bifurcations in the normal dispersion regime. WDM: wavelength division multiplex. EDF: Erbium-doped fiber. SMF: single-mode fiber. OC: optical coupler. PC: polarization controller. PD-ISO: polarization-dependent isolator. DCF: dispersion compensation fiber. PD: photo detector. OSA: optical spectrum analyzer.

*Dispersive Fourier transform measurement (DFT)*: The output pulse is linearly stretched by propagating in a long length of dispersion compensation fiber (DCF), so that the spectral intensity profile of the pulses is mapped into the time domain. The stretched signal is detected by a fast photodiode and recorded on a real-time digital oscilloscope. The DFT technique allows us to monitor the optical spectrum corresponding to each light pulse. The wavelength resolution of this setup is limited by the overall dispersion of the DCF and the bandwidth of the detection instruments.

In the anomalous dispersion regime, part of the laser output is directly sent to a 45-GHz photodiode to monitor the time-domain evolution on an oscilloscope with a 6 GHz bandwidth. Another part is sent through a 1250-m DCF, leading to an integrated dispersion of −128 ps/nm.

The spectral resolution of DFT is calculated to be 1.3 nm. The observed soliton spectra being typically more than 10 nm-wide, such resolution is suitable to follow the evolution of the spectrum roundtrip after roundtrip. In addition to these time-domain measurements, part of the output signal is also monitored on an optical spectrum analyzer (OSA) to obtain the time-averaged spectrum. Finally, autocorrelation traces of the output signal are also recorded on a time-averaged, second-order intensity autocorrelator.

In the normal dispersion regime, the pulses from the laser are linearly stretched in an 8.9-km long DCF, then detected with a 12.5 GHz photodetector and visualized on a high speed 8 GHz 25 GSa/s real-time oscilloscope. The total dispersion provided by the DCF is 459.2 $ps^2$ at 1550 nm. Consequently, the DFT measurement yields a spectral resolution of 0.2 nm.

**6. Simulation model**

To get insight into the physical mechanism of the period doubling observed in ultrafast lasers, we developed a detailed lumped simulation model of the experimental setup and carried out simulations. Pulse propagation in the fibers is modelled by using a generalized nonlinear Schrödinger equation (GNLSE):

$$\frac{\partial A}{\partial z} = \frac{g-\alpha}{2}A - i\frac{\beta_2}{2}\frac{\partial^2 A}{\partial t^2} + i\gamma|A|^2 A + \frac{g}{2\Omega_g^2}\frac{\partial^2 A}{\partial t^2}$$

where $A$ is the slowly-varying electric field in the time frame moving with the pulse. $g$ , $\alpha$ and $\beta_2$ are the gain, loss, and fiber dispersion, respectively. $\gamma$ is the fiber nonlinearity coefficient and $\Omega_g$ represents the gain bandwidth. Therefore, the first and the last terms on the right-hand side in the above equation account for the dissipative effects of gain saturation and spectral filtering arising in the doped fiber. The gain is described as $g = g_0 exp(-E_p/E_{sat})$, where $g_0$ is the small signal gain, $E_{sat}$ is the fiber saturation energy and $E_p$ denotes the instantaneous pulse energy in the cavity. For simplicity, the SA is modelled by an instantaneous nonlinear transfer function that writes: $T = 1 - \left[a_n + \frac{a_0}{1+\frac{I(t)}{I_{sat}}}\right]$, where $a_n$ is the linear loss, $a_0$ is the saturable modulation depth, $I_{sat}$ is the saturation power and $I(t)$ denotes the instantaneous power in the cavity. The laser cavity components and their parameters are listed in the tables below. In experiments, the artificial SA via NPE in the normal dispersion regime also adds a spectral filtering effect, which we modelled by a 6 nm Gaussian filter. The initial condition is a weak hyperbolic secant (sech) pulse mixed with a white noise field. We just vary the small signal gain parameter in the simulations to achieve the different lasing states, the rest of parameters being fixed as indicated in the Tables below. To simulate the bifurcation curve, the solution of the current value of gain serves as the initial condition for the next gain value.

**Table 1.** Parameters for anomalous dispersion cavity

| Components | EDF | SMF1 | OC | SMF2 | SA | SMF3 |
|---|---|---|---|---|---|---|
| | $\beta_2$ =13.5 [ps$^2$/km]<br><br>$\gamma$ = 1.69 [W$^{-1}$km$^{-1}$]<br><br>$\Omega_g$ = 80 [nm]<br><br>$E_{sat}$ = 105 [pJ]<br><br>L = 0.5 [m] | $\beta_2$ = -22.87 [ps$^2$/km]<br><br>$\gamma$ = 1.3 [W$^{-1}$km$^{-1}$]<br><br>$L$ = 0.5 [m] | $T$ [a] = 0.85 | $\beta_2$ = -22.87 [ps$^2$/km]<br><br>$\gamma$ = 1.3 [W$^{-1}$km$^{-1}$]<br><br>$L$ = 3.5 [m] | $a_n$ [b] = 0.43<br><br>$a_0$ [c] = 0.19<br><br>$P_{sat}$ = 100 [W] | $\beta_2$ = -22.87 [ps$^2$/km]<br><br>$\gamma$ = 1.3 [W$^{-1}$km$^{-1}$]<br><br>$L$ = 1.8 [m] |

a)( $T$ : coupling ratio of optical coupler); b)( $a_n$: linear absorption); c)( $a_0$: modulation depth)

**Table 2.** Parameters for normal dispersion cavity

| Components | EDF | SMF1 | OC | SA | Filter | SMF2 | DCF |
|---|---|---|---|---|---|---|---|
| | $\beta_2$ = 61.2 [ps$^2$/km]<br><br>$\gamma$ = 3.144 [W$^{-1}$km$^{-1}$]<br><br>$\Omega_g$ = 80 [nm]<br><br>$E_{sat}$ = 150 [pJ]<br><br>$L$ = 1 [m] | $\beta_2$ = -22.87 [ps$^2$/km]<br><br>$\gamma$ = 1.3 [W$^{-1}$km$^{-1}$]<br><br>$L$ = 5.0 [m] | $T$ = 0.9 | $a_n$ = 0.53<br><br>$a_0$ = 0.43<br><br>$P_{sat}$ = 200 [W] | $\lambda_c$ = 1560 [nm]<br><br>$\Delta\lambda$= 6 [nm]<br><br>Gaussian profile | $\beta_2$ = -22.87 [ps$^2$/km]<br><br>$\gamma$ = 1.3 [W$^{-1}$km$^{-1}$]<br><br>$L$ = 0.35 [m] | $\beta_2$ = 56.1 [ps$^2$/km]<br><br>$\gamma$ = 2.113 [W$^{-1}$km$^{-1}$]<br><br>$L$ = 3.6 [m] |

The consistence of the simulation results with experimental observations shows that a GNLSE-based lumped model is a simple while universal tool to describe complex nonlinear dynamics in dissipative systems.

**Supporting Information**

Supporting Information is available from the Wiley Online Library or from the author.

**Acknowledgements**

European Commission Marie Curie Individual Fellowship (891017); Jiangsu Shuangchuang Outstanding Doctor Talents Support Program (CZ1060619002); Natural Science Foundation of Jiangsu Province (BK20180742); National Natural Science Foundation of China (91950105 and 62175116).

Part of this work was funded by the ANR CoMuSim (ANR-17-CE24-0010-01) and benefited from the facilities of the SMARTLIGHT platform in Bourgogne Franche-Comté (EQUIPEX+ contract ANR-21-ESRE-0040) and the French "Investissement d'Avenir" program ISITE-BFC (contract ANR-15-IDEX-0003).

**Conflict of Interest**

The authors declare no conflict of interest.

**Author Contributions**

Zhiqiang Wang and Aurélien Coillet contributed equally to this work. Zhiqiang Wang, Saïd Hamdi and Aurélien Coillet performed the experiments. Zhiqiang Wang carried out the simulations and analyzed the data. Zhiqiang Wang and Philippe Grelu wrote the main draft. Zhiqiang Wang, Aurélien Coillet and Philippe Grelu discussed the results. All authors commented on the manuscript. Zuxing Zhang and Philippe Grelu supervised the project.

**References**

[1] C. K. Aidun, and E.-J. Ding, *Phys. Fluids* **15**, 1612–1621 (2003).

[2] A. G. Tomboulides, G. S. Triantafyllou, and G. E. Karniadakis, *Phys. Fluids A Fluid Dyn.* **4**, 1329–1332 (1992).

[3] O. Tzuk, S. R. Ujjwal, C. Fernandez-Oto, M. Seifan, and E. Meron, *Sci. Rep.* **9**, 1–10 (2019).
[4] J. Hansen, M. Sato, and R. Ruedy, *Proc. Natl. Acad. Sci.* **109**, E2415–E2423 (2012).
[5] M. R. Guevara, L. Glass, and A. Shrier, *Science (80-. ).* **214**, 1350–1353 (1981).
[6] T. Quail, A. Shrier, and L. Glass, *Proc. Natl. Acad. Sci.* **112**, 9358–9363 (2015).
[7] J. L. Aron, and I. B. Schwartz, *J. Theor. Biol.* **110**, 665–679 (1984).
[8] N. Akhmediev, J. M. Soto-Crespo, and G. Town, *Phys. Rev. E* **63**, 56602 (2001).
[9] G. Sucha, S. R. Bolton, S. Weiss, and D. S. Chemla, *Opt. Lett.* **20**, 1794–1796 (1995).
[10] D. Hennig, *Phys. Rev. E* **59**, 1637 (1999).
[11] S. Coen, M. Haelterman, P. Emplit, L. Delage, L. M. Simohamed, and F. Reynaud, *JOSA B* **15**, 2283–2293 (1998).
[12] D. Côté, and H. M. van Driel, *Opt. Lett.* **23**, 715–717 (1998).
[13] X. Xiao, Y. Ding, S. Fan, C. Yang, and X. Zhang, *ArXiv Prepr. ArXiv2101.00149* (2021).
[14] B. Cao, K. Zhao, C. Gao, X. Xiao, C. Bao, and C. Yang, *ArXiv Prepr. ArXiv2109.00989* (2021).
[15] P. Grelu, and N. Akhmediev, *Nat. Photonics* **6**, 84–92 (2012).
[16] J. M. Soto-Crespo, M. Grapinet, P. Grelu, and N. Akhmediev, *Phys. Rev. E* **70**, 66612 (2004).
[17] M. J. Lederer, B. Luther-Davies, H. H. Tan, C. Jagadish, N. N. Akhmediev, and J. M. Soto-Crespo, *JOSA B* **16**, 895–904 (1999).
[18] Z. Q. Wang, K. Nithyanandan, A. Coillet, P. Tchofo-Dinda, and P. Grelu, *Nat. Commun.* **10**, 1–11 (2019).
[19] K. Krupa, K. Nithyanandan, U. Andral, P. Tchofo-Dinda, and P. Grelu, *Phys. Rev. Lett.* **118**, 243901 (2017).
[20] J. Peng, and H. Zeng, *Laser Photon. Rev.* **12**, 1800009 (2018).
[21] X. Liu, X. Yao, and Y. Cui, *Phys. Rev. Lett.* **121**, 23905 (2018).
[22] J. M. Soto-Crespo, N. Akhmediev, and A. Ankiewicz, *Phys. Rev. Lett.* **85**, 2937 (2000).
[23] E. N. Tsoy, and N. Akhmediev, *Phys. Lett. A* **343**, 417–422 (2005).
[24] E. N. Tsoy, A. Ankiewicz, and N. Akhmediev, *Phys. Rev. E* **73**, 36621 (2006).
[25] J. Peng, S. Boscolo, Z. Zhao, and H. Zeng, *Sci. Adv.* **5**, eaax1110 (2019).
[26] J. Peng, Z. Zhao, S. Boscolo, C. Finot, S. Sugavanam, D. V Churkin, and H. Zeng, *Laser Photon. Rev.* **15**, 2000132 (2021).

[27] K. Krupa, T. M. Kardas, and Y. Stepanenko, in: Real-Time Meas. Rogue Phenomena, Single-Shot Appl. VI: (International Society for Optics and Photonics, 2021), p. 116710E.
[28] M. Liu, Z. Wei, H. Li, T. Li, A. Luo, W. Xu, and Z. Luo, *Laser Photon. Rev.* **14**, 1900317 (2020).
[29] J. Peng, and H. Zeng, *Phys. Rev. Appl.* **11**, 44068 (2019).
[30] J. Peng, and H. Zeng, *Phys. Rev. Appl.* **12**, 34052 (2019).
[31] J. M. Dudley, F. Dias, M. Erkintalo, and G. Genty, *Nat. Photonics* **8**, 755–764 (2014).
[32] B. Kibler, J. Fatome, C. Finot, G. Millot, F. Dias, G. Genty, N. Akhmediev, and J. M. Dudley, *Nat. Phys.* **6**, 790–795 (2010).
[33] M. Yu, J. K. Jang, Y. Okawachi, A. G. Griffith, K. Luke, S. A. Miller, X. Ji, M. Lipson, and A. L. Gaeta, *Nat. Commun.* **8**, 1–7 (2017).
[34] E. Lucas, M. Karpov, H. Guo, M. L. Gorodetsky, and T. J. Kippenberg, *Nat. Commun.* **8**, 1–11 (2017).
[35] S. T. Cundiff, J. M. Soto-Crespo, and N. Akhmediev, *Phys. Rev. Lett.* **88**, 73903 (2002).
[36] Z. Wang, K. Nithyanandan, A. Coillet, P. Tchofo-Dinda, and P. Grelu, *Phys. Rev. Res.* **2**, 13101 (2020).
[37] J. M. Soto-Crespo, N. N. Akhmediev, V. V Afanasjev, and S. Wabnitz, *Phys. Rev. E* **55**, 4783 (1997).
[38] F. W. Wise, A. Chong, and W. H. Renninger, *Laser Photon. Rev.* **2**, 58–73 (2008).
[39] A. Chong, J. Buckley, W. Renninger, and F. Wise, *Opt. Express* **14**, 10095–10100 (2006).
[40] F. Ö. Ilday, J. R. Buckley, W. G. Clark, and F. W. Wise, *Phys. Rev. Lett.* **92**, 213902 (2004).
[41] R. J. Deissler, and H. R. Brand, *Phys. Rev. Lett.* **72**, 478 (1994).
[42] W. Chang, J. M. Soto-Crespo, P. Vouzas, and N. Akhmediev, *Phys. Rev. E* **92**, 22926 (2015).
[43] K. Tamura, C. R. Doerr, H. A. Haus, and E. P. Ippen, *IEEE Photonics Technol. Lett.* **6**, 697–699 (1994).
[44] L. Luo, T. J. Tee, and P. L. Chu, *JOSA B* **15**, 972–978 (1998).
[45] G. Sucha, D. S. Chemla, and S. R. Bolton, *JOSA B* **15**, 2847–2853 (1998).
[46] H.-J. Chen, Y.-J. Tan, J.-G. Long, W.-C. Chen, W.-Y. Hong, H. Cui, A.-P. Luo, Z.-C. Luo, and W.-C. Xu, *Opt. Express* **27**, 28507–28522 (2019).
[47] Y. Du, Z. Xu, and X. Shu, *Opt. Lett.* **43**, 3602–3605 (2018).

[48] J. Chen, X. Zhao, T. Li, J. Yang, J. Liu, and Z. Zheng, *Opt. Express* **28**, 14127–14133 (2020).

[49] Y. Cui, Y. Zhang, Y. Song, L. Huang, L. Tong, J. Qiu, and X. Liu, *Laser Photon. Rev.* **15**, 2000216 (2021).

[50] X. Yao, L. Li, A. Komarov, M. Klimczak, D. Tang, D. Shen, L. Su, and L. Zhao, *Opt. Express* **28**, 9802–9810 (2020).

[51] K. Goda, and B. Jalali, *Nat. Photonics* **7**, 102–112 (2013).

[52] J. Desbois, F. Gires, and P. Turnois, *IEEE J. Quantum Electron.* **9**, 213–218 (1973).

[53] K. Tamura, E. P. Ippen, H. A. Haus, and L. E. Nelson, *Opt. Lett.* **18**, 1080–1082 (1993).

[54] S. Chouli, J. M. Soto-Crespo, and P. Grelu, *Opt. Express* **19**, 2959–2964 (2011).

[55] D. C. Cole, and S. B. Papp, *Phys. Rev. Lett.* **123**, 173904 (2019).

[56] T. Xian, L. Zhan, W. Wang, and W. Zhang, *Phys. Rev. Lett.* **125**, 163901 (2020).

[57] A. Chong, W. H. Renninger, and F. W. Wise, *JOSA B* **25**, 140–148 (2008).

[58] C. Lecaplain, J. M. Soto-Crespo, P. Grelu, and C. Conti, *Opt. Lett.* **39**, 263–266 (2014).

[59] H. A. Haus, K. Tamura, L. E. Nelson, and E. P. Ippen, *IEEE J. Quantum Electron.* **31**, 591–598 (1995).

[60] T. Herr, V. Brasch, J. D. Jost, C. Y. Wang, N. M. Kondratiev, M. L. Gorodetsky, and T. J. Kippenberg, *Nat. Photonics* **8**, 145–152 (2014).

[61] H. Guo, E. Lucas, M. H. P. Pfeiffer, M. Karpov, M. Anderson, J. Liu, M. Geiselmann, J. D. Jost, and T. J. Kippenberg, *Phys. Rev. X* **7**, 41055 (2017).

[62] A. Coillet, and Y. K. Chembo, *Chaos An Interdiscip. J. Nonlinear Sci.* **24**, 13113 (2014).

ToC: The periodical doubling of dissipative solitons in ultrafast lasers under different dispersion regime has been observed, demonstrating the universality of periodical doubling in nonlinear systems, and revealing the different intertwined bifurcations and the entrainment of new pulsating frequencies in the period doubling bifurcations as well as the synchronization of the oscillating periodicity with integer multiples of the cavity roundtrip.

*Zhiqiang Wang[1,3], Aurélien Coillet[2], Saïd Hamdi[2], Zuxing Zhang[3, *] and Philippe Grelu[2, *]*

**Spectral pulsations of dissipative solitons in ultrafast fiber lasers: period doubling and beyond**

ToC figure:

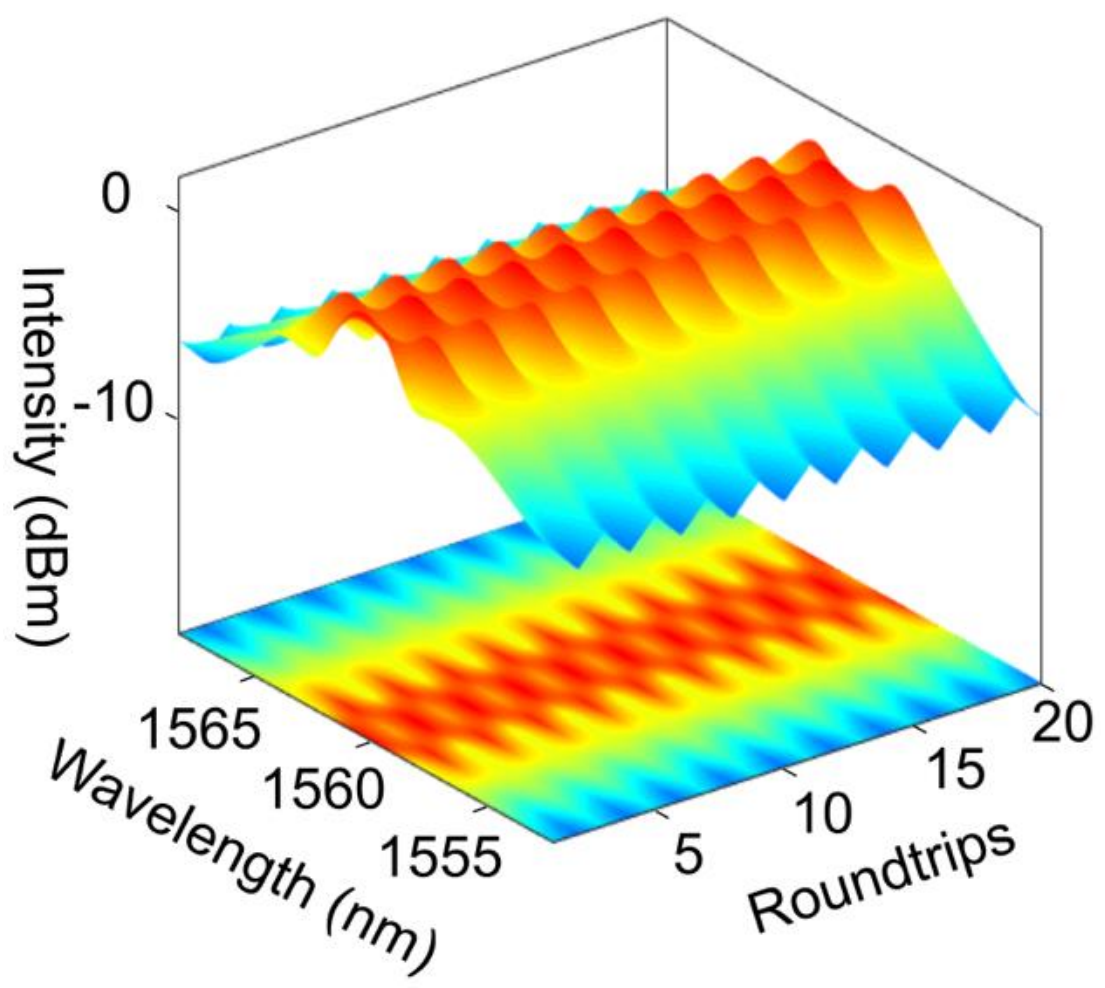

**Supporting Information**

**Spectral pulsations of dissipative solitons in ultrafast fiber lasers: period doubling and beyond**

*Zhiqiang Wang[1,3], Aurélien Coillet[2], Saïd Hamdi[2], Zuxing Zhang[3, *] and Philippe Grelu[2, *]*

**S1. Spectral pulsations of dissipative solitons in ultrafast lasers: diversity and universality**

*S1.1 Period-2 dynamics with enhanced spectral alterations (anomalous dispersion regime)*

Below, we provide another experimental recording of period-2 spectral pulsations, which displays the alternation of more pronounced spectral features, with an oscillating double-hump in the central spectral region. This example, obtained in the anomalous dispersion regime, is another illustration of "invisible pulsations", namely associated with unnoticeable variations of the pulse energy over consecutive roundtrips.

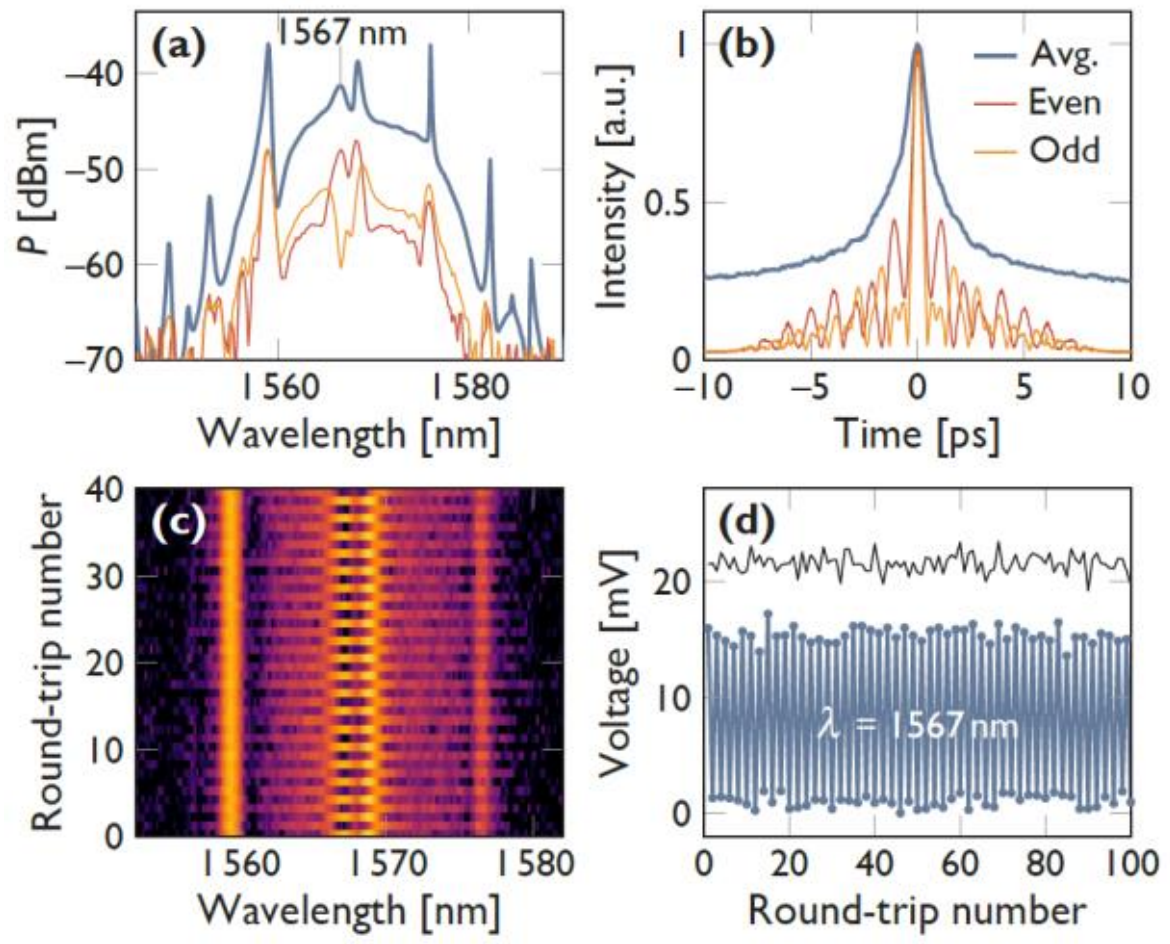

**Figure S1.** Period doubling of dissipative solitons in anomalous dispersion ultrafast fiber laser. (a) Time- averaged spectrum (blue curve) and DFT spectra for two successive roundtrips. (b) Time-averaged autocorrelation trace of the pulse train (blue curve) and first-order AC trace obtained by Fourier transforming the single-shot DFT spectra. (c) Evolution map of the DFT measurement showing the stable spectrum oscillations. (d) Evolution with the round-trip number of the intensity of the pulse (thin black) and of the 1567 nm component using the DFT measurements. The oscillations of the intensity at 1567 nm are stable all over the long recording time (> 4000 roundtrips).

*S1.2 Period-3 pulsations of a dissipative soliton molecule (anomalous dispersion regime)*

Here, we show the existence of a period-3 pulsation in the case of an optical soliton molecule, self-generated from the ultrafast fiber laser. Not as general as the period-2 bifurcation, period-3 is nevertheless a known possibility of the complex dissipative soliton dynamics.[16] Here, it applies to the multi-pulsing instability that has resulted from the increase of the pump power, which led to the generation of two dissipative soliton pulses and their self-assembly into a soliton molecule. In the present situation, we verify that the pulsations do not correspond to internal motions within the soliton molecule found elsewhere, such as the oscillations of the relative separation and phase between the two solitons. The latter parameters remain constant, as attested by the averaged recording of the optical spectrum, which displays highly contrasted spectral fringes. Instead, the pulsation concerns the spectral intensity affecting both pulses in a similar way.

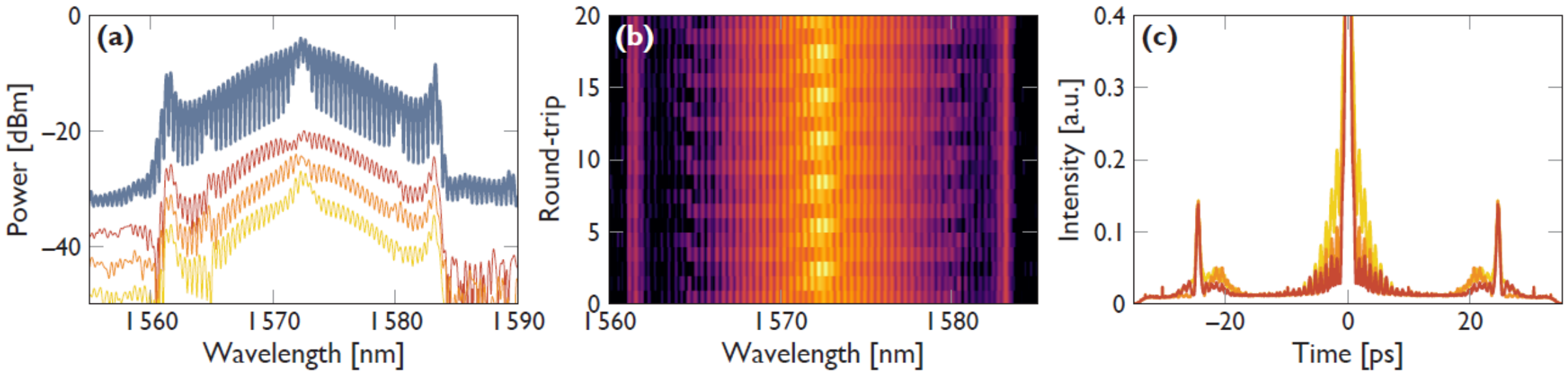

**Figure S2.** Period-3 spectral pulsations of a dissipative soliton molecule. (a) Average spectrum from the OSA (dark blue) and 3 consecutive spectra from the DFT measurements. (b) DFT map showing the period 3 evolution of the soliton molecule. (c) single-shot and averaged first-order autocorrelation traces obtained from the Fourier transform of the DFT signal.

*S1.3 Period-4 synchronization of dissipative solitons in a Tm-doped mode-locked laser oscillator*

To highlight the universality of spectral pulsations, we present experimental observations performed in a fiber laser operating around the wavelength of 1.9 μm. The laser setup is based on a fiber ring cavity that incorporates thulium-doped fiber pumped at 1565 nm that provides gain in the 1.85-2.0 μm wavelength region. Due to its moderate absorption (9 dB/m at 1180 nm), the thulium fiber is 3-m long, and a pumping power of 500 mW is required for stable soliton generation. The effective saturable absorber function is performed by the nonlinear evolution of the polarization in the SMF fiber that constitutes the remaining part of the cavity, leading to an anomalous overall dispersion. In the Figs. S3 (a, b, c) below, we display the recording of stationary period-4 spectral pulsations. Moreover, we show the synchronization of the spectral pulsation to an exact 4 roundtrip periods in Figs. S3 (d), obtained in the course of increasing the pump power from 500 mW to 600 mW, and recorded over 4000 cavity roundtrips.

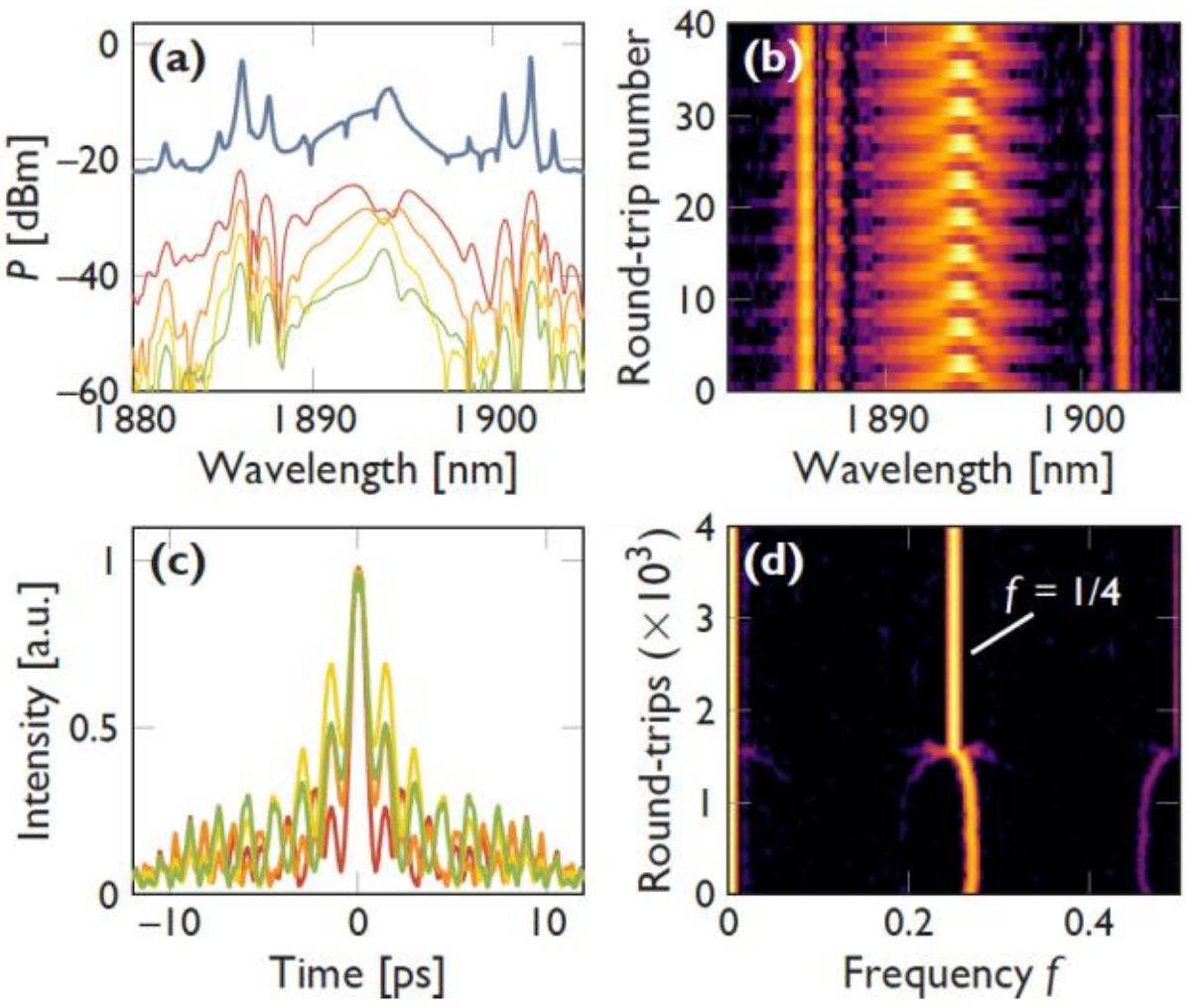

**Figure S3.** Period-4 solitons in a thulium fiber laser emitting around 1.9 µm. The fiber laser architecture is similar to the one shown in figure 10, with a thulium-doped fiber replacing the erbium-doped fiber. (a) Average spectrum from the OSA and 4 different spectra from the DFT measurements. (b) DFT map showing the period 4 evolution of the soliton. (c) First-order autocorrelation traces obtained from the Fourier transform of the DFT signal. (d) Spectrogram of the intensity at 1894 nm, revealing a continuous variation of the period until it locks onto a value corresponding to 4 times the round-trip.

### S2. Bifurcations in normal dispersion regime

Fig. S4 shows the simulation results of the bifurcation of the pulse peak power (Fig. 5S(a)) and pulse energy (Fig. 5S(b)) as a function of the gain parameter in an ultrafast laser in the normal dispersion regime. The gain ($g_0$) is varying from 4 m$^{-1}$ to 17 m$^{-1}$ with a step of 0.5 m$^{-1}$. The first bifurcation from period one to period doubling occurs at $g_0$= 11.5 m$^{-1}$. Different than the smooth transition for the anomalous dispersion cases shown in Fig. 5(i) in the main text, the occurrence of the period-2 solitons comes with a significant pulse compression process, which is reflected by the sudden increase in the pulse peak power and slowly growth of the pulse energy. The pulse compression process is attributed to the excess of the nonlinearity under high pump power in the normal dispersion regime. Further increasing the gain leads to the cascaded bifurcations of period-2 and period-N operations. The zoom-in of the pulse energy evolution in the insert in Fig. S5(b) shows the period-2 bifurcation of the pulse energy, which confirms the visible pulsating solitons observed in the experiments (See Fig. 6 in the main text).

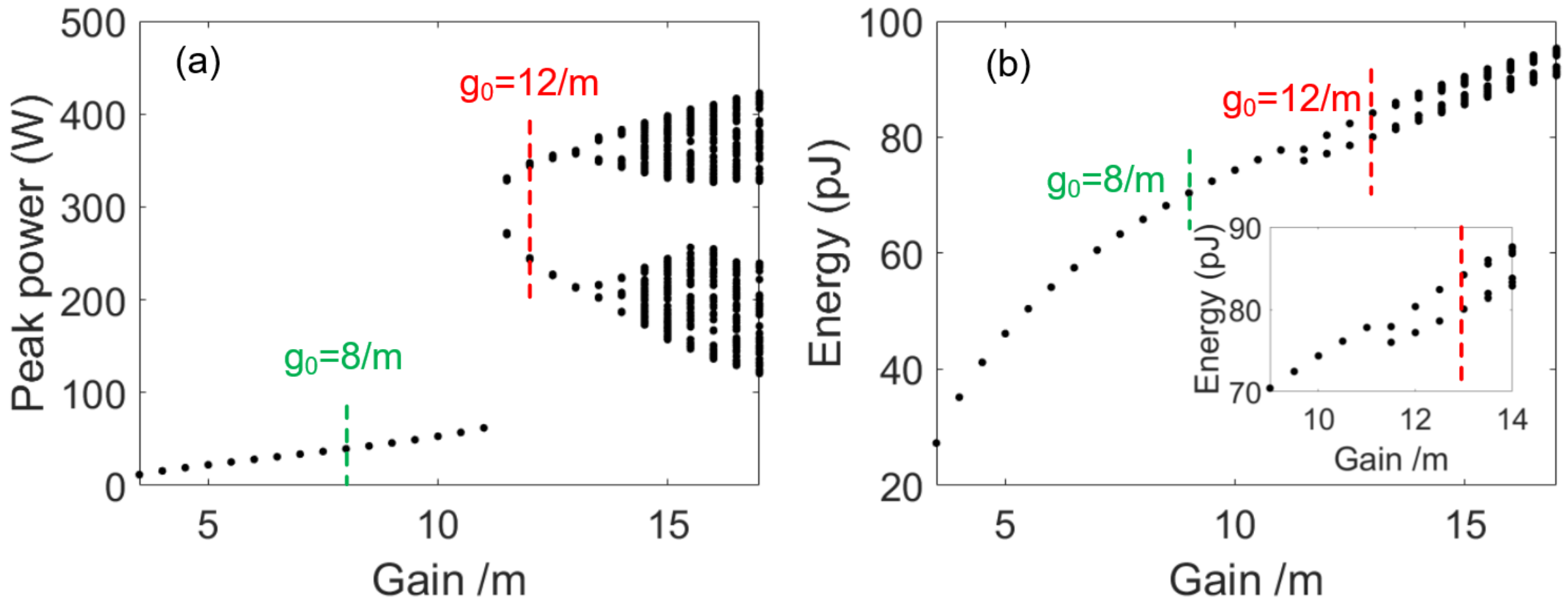

**Figure S4.** Simulation results of the evolution of intracavity pulse peak power and pulse energy as a function of $g_0$ showing the sequence of from period-1 to period-2 and then to period-N bifurcations. (a) Pulse peak power. (b) Pulse energy. The first period doubling bifurcation occurs at $g_0$ = 11.5 m$^{-1}$. As $g_0$ increases from 11.5 m$^{-1}$ to 17 m$^{-1}$, the period orbits undergo a sequence of period doubling bifurcation with orbits tending to ∞.